\documentclass[preprint]{JASA}

\usepackage{amsmath}
\usepackage{mathtools}
\usepackage{hhline}
\usepackage{multirow}
\usepackage{url}
\usepackage{booktabs}
\usepackage[caption=false,font=footnotesize]{subfig}
\usepackage{array}
\usepackage{enumitem}
\usepackage{dirtytalk}
\usepackage{siunitx}
\usepackage{float}
\usepackage{mathrsfs}
\usepackage{upgreek}

\begin{document}

\title{Spatially Selective Active Noise Control Systems}
\author{Tong Xiao}
\email{Tong.Xiao@uts.edu.au}
\affiliation{Centre for Audio, Acoustics and Vibration, University of Technology Sydney, NSW 2007, Australia}

\author{Buye Xu}
\email{xub@meta.com}
\affiliation{Meta Reality Labs Research, Redmond, WA 98052, USA}
 
\author{Chuming Zhao}
\email{chumingz@meta.com}
\affiliation{Meta Reality Labs Research, Redmond, WA 98052, USA}


\date{\today}

\begin{abstract}
Active noise control (ANC) systems are commonly designed to achieve maximal sound reduction regardless of the incident direction of the sound. When desired sound is present, the state-of-the-art methods add a separate system to reconstruct it. This can result in distortion and latency. In this work, we propose a multi-channel ANC system that only reduces sound from undesired directions, and the system truly preserves the desired sound instead of reproducing it. The proposed algorithm imposes a spatial constraint on the hybrid ANC cost function to achieve spatial selectivity. Based on a six-channel microphone array on a pair of augmented eyeglasses, results show that the system minimized only noise coming from undesired directions. The control performance could be maintained even when the array was heavily perturbed. The proposed algorithm was also compared with the existing methods in the literature. Not only did the proposed system provide better noise reduction, but it also required much less effort. The binaural localization cues were not needed to be reconstructed since the system preserved the physical sound wave from the desired source.
\end{abstract}


\maketitle

\section{Introduction}
Active noise control (ANC) systems have seen many significant advancements over the past few decades. Notably, many personal ANC headphones, aiming to eliminate the unwanted noise for users, have emerged and gained much success in the market due to their excellent performance and robustness~\citep{Chang2016Listening}. Other ANC systems, such as ANC headrests~\citep{elliott2018head,xiao2020} and ANC windows~\citep{wang2017controlling,lam2020active}, have also seen significant improvements over the years. These ANC systems have been designed to attenuate all the sound in the environment. One emerging application is using ANC to enhance face-to-face conversations in a noisy environment, e.g., a cocktail-party scenario, where one hopes to 
minimize the surrounding noises while still maintaining the conversations in front of the user. An intelligent ANC system should separate and categorize sounds coming from various directions. The desired sound should be maintained for the user while noises coming from other directions are minimized.

The goal of this work is to develop a spatially selective ANC system that preserves sound coming from the desired directions while reducing noise from other directions. The filtered-reference least-mean-square (FxLMS) algorithm is commonly used in these systems for adaptive control due to its simplicity and robustness~\citep{Kuo1996a,Elliott2000,Hansen2012}. Any signals observed by the reference microphones (or error microphones in the feedback systems) will be fully controlled regardless of the residual noise spectrum at the error microphones. So far, most effort in the ANC systems has been devoted to improving the noise reduction level throughout the spectrum as much as possible.

In recent years, some studies have considered the spatial aspect of personal ANC systems. Studies~\citep{Rafaely2002combined,Zhang2014causality,Liebich2018DOA,Cheer2019} have showed that the performance in ANC headphones and earphones was affected by the time advance between the reference and the error microphones in the feedforward configurations. The ANC performance was the best when the time advance was the most significant. However, these systems were studied to further improve the control performance when dealing with direction- and/or time-varying noises. They were not designed for spatial noise selection. 

There are systems, which were not intentionally designed, that can be modified to achieve spatial selection functionality. For example, the coherence-based selection method~\citep{Shen2021wireless} can potentially be used to determine the direction of arrival (DOA) of the noise, which can then be isolated. The selective fixed-filter method~\citep{Shi2020FFselective,SHI2022Selective} and the deep ANC method~\citep{zhang2021deep} can be extended to selecting spatial filters to control noise coming from certain DOAs. However, these systems still possess issues. The coherence-based selection method requires the input signals to be distinguished enough to offer differences in the coherence functions. Thus, it may be limited to certain spread-out array configurations, e.g., an array distributed in a room. The selective fixed-filter method and the deep ANC method require the system to be pre-trained to obtain the optimal filters in advance. These two methods also require the spatial information to be acquired elsewhere and thus resulting in additional computations. These systems are not yet the best solutions for a spatially selective ANC system.

To achieve spatial selectivity, it is common to use the beamforming technique. Beamforming is a well-established method of designing spatio-temporal filters for array processing~\citep{vanveen1988,VanTrees2002,Doclo2015}. For personal devices, a certain number of microphones can be used to differentiate sound from different directions. Hearing aids typically employ it to improve speech intelligibility for the hearing impaired. Studies have incorporated the ANC functionality to control the leakage sound in open-fitting systems, which improves the insertion signal-to-noise-ratio (SNR) gain of the multi-channel Wiener filter (MWF)~\citep{Serizel2010,Dalga2011}. \citet{Serizel2010} used the feedforward ANC to control the leakage noise, while \citet{Dalga2011} further improved the result by using the hybrid control due to its better ANC performance. However, the desired sound in the leakage is canceled (together with the noise) and then added back to the error signal with a delay. Recent work in~\citep{Patel2020} proposed a structure for a pair of hear-through ANC headphones. The microphones were grouped for different purposes independently, i.e., ANC and beamforming, and the desired sound from the beamformer was injected into the secondary sources. 

The studies from above can be improved regarding the following three aspects:
\begin{enumerate} 
    \item \textbf{Control effort} Current studies require canceling both noise and the desired sound first, and then reproducing this desired sound again. Thus, the control effort can be significant.
    \vspace{-2mm}
    \item \textbf{ANC performance} When an adaptive ANC system needs to reduce both noise and the desired sound, the attenuation of the noise may be degraded because the desired sounds, e.g., speech and music, are often non-stationary, which can drive the adaptive system to sub-optimal states. 
    \vspace{-2mm}
    \item \textbf{Distortion to the desired sound} When the desired sound needs to be reconstructed and reproduced acoustically, it introduces latency and is prone to different types of distortions, e.g., undesired frequency shaping or binaural cue distortion for ear-level devices. 
\end{enumerate}
These aspects lead to the question, ``Instead of minimizing the noise plus the desired sound, and then \replaced{add}{adding} the desired sound once again, can the system control only the noise but not the desired sound?''

This article, based on the previous work~\citep{Xu2019}, proposes a spatially selective ANC system that only controls the unwanted noise while leaving the desired physical sound unaltered. The proposed algorithm is derived based on the hybrid ANC architecture when both the reference and the error signals can be used as inputs for both ANC and the spatial constraint at the same time to optimize the performance. 
The constraint vector describing the frequency response of the signal from the desired DOA at the error microphone will make sure the desired signal component is physically preserved rather than reconstructed to avoid the aforementioned three issues.



\section{Proposed Spatially Selective ANC Systems} \label{Section:Proposed_algorithm}
\subsection{Signal definition and cost function} \label{Section:Proposed_signal_def_cost_function}
\begin{figure}[t]
        \centering
        \includegraphics[width=2.6in]{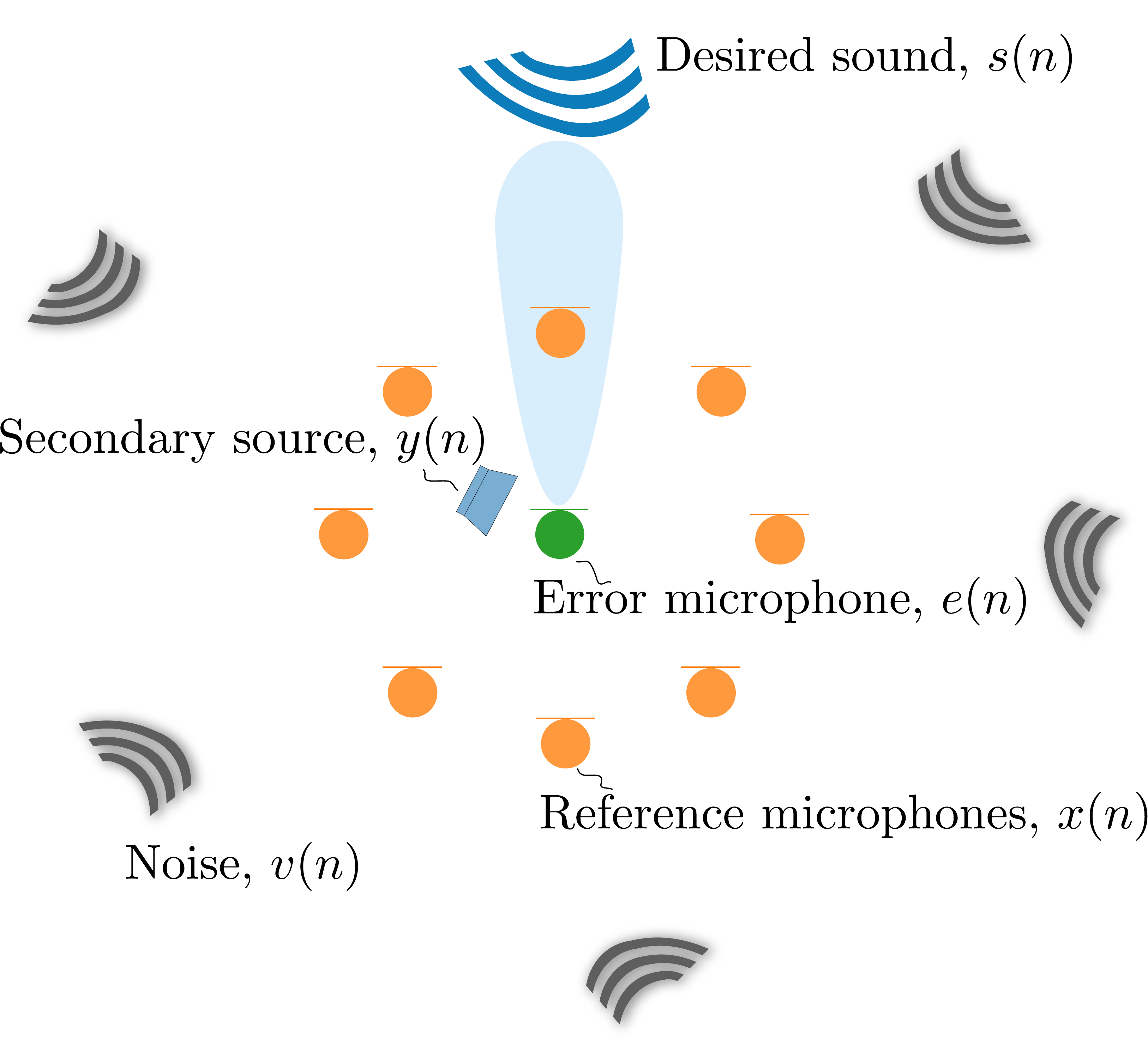}
        \caption{Concept diagram of a spatially selective ANC system \textit{preserving} the desired sound from one direction and using a secondary source to control noise from other directions at the error microphone (one shown here for demonstration).}
        \label{fig:1}
    \end{figure}

Without losing generality, we assume there is one desired sound source and multiple noise sources in the scene (see Fig.~\ref{fig:1}).
The ANC system contains a total of $K$ microphones, one of which serves as an error microphone. One error microphone with one secondary source is used in the derivation hereinafter for notation brevity, but it can be easily extended to cases with multiple sources and/or microphones. The disturbance signal, $d(n)$, at the error microphone with ANC disabled includes two components,
\begin{equation}
    d(n) = s(n) + v(n),
    \label{eq:d_s_v}
\end{equation}
where $s(n)$ and $v(n)$ are the desired signal and the noise signal, respectively, and $n$ denotes the time index. \added{We assume that the two signals come from different locations and there is one desired source for now.}

We write the error signal $e(n)$ of the hybrid ANC system in \textit{matrix multiplication form} as~\citep{Hansen2012}
\begin{subequations}
\begin{align}
    e(n) &= d(n) + \mathbf{w}^\text{T}{\mathbf{G}}^\text{T}{\mathbf{x}}(n)
    \\
    &=  \tilde{\delta}^\text{T}{\mathbf{x}}(n) + \mathbf{w}^\text{T}{\mathbf{G}}^\text{T}{\mathbf{x}}(n)
    \\
    &= \mathbf{u}^\text{T} {\mathbf{x}}(n),
\end{align}
\label{eq:e=d+wGx}
\end{subequations}
where 
\begin{subequations}
\begin{align}
    \hspace{-0.5mm} \mathbf{w} \in \mathbb{R}^{KL} &= \left[\mathbf{w}^\text{T}_1 \ \mathbf{w}^\text{T}_2 \ \dots \ \mathbf{w}^\text{T}_K \right]^\text{T},
\\ \notag
\\[-12pt]
    \hspace{-0.5mm} \mathbf{w}_k \in \mathbb{R}^{L} &= \left[{w_k}_0 \ {w_k}_1 \ \dots \ {w_k}_{(L-1)} \right]^\text{T}, 
\\ \notag
\\[-12pt]
    \hspace{-0.5mm} {\mathbf{G}} \in \mathbb{R}^{KL \times KL} 
    &= \text{diag}\left(\hat{\mathbf{G}} \ \hat{\mathbf{G}} \ \dots \ \hat{\mathbf{G}} \right),
    \label{eq:G_tilde_multi}
\\
    \hspace{-0.5mm} \hat{\mathbf{G}} \in \mathbb{R}^{L \times L}
    &=
    \left[\begin{array}{ccccc}
    \hat{g}_{0} & 0 & 0 & \cdots & 0 \\
    \hat{g}_{1} & \hat{g}_{0} & 0 & \cdots & 0 \\
    \hat{g}_{2} & \hat{g}_{1} & \hat{g}_{0} & \cdots & 0 \\
    \vdots & \vdots & \vdots & \ddots & \vdots \\
    \hat{g}_{L-1} & \hat{g}_{L-2} & \hat{g}_{L-3} & \cdots & \hat{g}_{0}
    \end{array}\right] ,
    \label{eq:G_hat}
\\
    \hspace{-0.5mm} {\mathbf{x}}(n) \in \mathbb{R}^{KL} 
    &= 
    \left[\mathbf{x}_{1}^\text{T}(n) \ \dots \ \mathbf{x}_{K-1}^\text{T}(n) \ \hat{\mathbf{d}}^\text{T}(n) \right]^\text{T},
    \label{eq:x_tilde_multi}
\\
    \hspace{-0.5mm} \mathbf{x}_k(n) \in \mathbb{R}^{L} 
    &= 
    \left[ x_k(n) \ x_k(n-1) \ \dots \ x_k(n-L+1) \right]^\text{T},
\\ 
    \hspace{-0.5mm} \hat{\mathbf{d}}(n) \in \mathbb{R}^{L} 
    &= 
    \left[ \hat{d}(n) \ \hat{d}(n-1) \ \dots \ \hat{d}(n-L+1) \right]^\text{T},
\\
    \hspace{-0.5mm} \tilde{\delta} \in \mathbb{R}^{KL} 
    &= 
    \left[\begin{array}{ccccc}
    \mathbf{0}^\text{T} & \ldots & \mathbf{0}^\text{T} & \delta^\text{T}
    \end{array}\right]^\text{T},
    \label{eq:delta_tilde_multi}
\\ \notag
\\[-12pt]
    \hspace{-0.5mm} \mathbf{u} \in \mathbb{R}^{KL}
    &=
    \tilde{\delta} + {\mathbf{G}} \mathbf{w},
    \label{eq:u_equals_w}
\end{align}
\end{subequations}
where $\mathbf{w}_k$ is the control filter with $L$ length for the $k\text{-th}$ channel $(k = 1, 2, \dots, K)$. Superscript $(\cdot )^\text{T}$ denotes the transpose, and hat accent $\hat{ \; }$ represents the estimated value. $\hat{\mathbf{G}}$ is the Toeplitz matrix of $\hat{\mathbf{g}}$, which is the estimation of the secondary path impulse response. We assume that it agrees well with the actual one, i.e., $\hat{\mathbf{g}} = \left[\hat{g}_0\ \hat{g}_1\ \dots\ \hat{g}_{L-1}\right]^\text{T} \approx \mathbf{g}$. This assumption holds in many situations and will be made hereinafter. $\delta = [1 \ 0 \ \dots \ 0]^\text{T}$ is the Dirac delta function. 
$\hat{\mathbf{d}}(n)$ is the vector of the estimated past values of the disturbance signal $\mathbf{d}(n)$.
At the time $n+1$, and after when the ANC is enabled, it is recovered from the error signal $e(n)$ as
\begin{equation}
\hat{d}(n+1)=e(n)-\hat{\mathbf{g}}^\text{T} \mathbf{y}(n),
\label{eq:d_hat}
\end{equation}
where $y(n)$ is the secondary source signal~\citep{Kuo1996a}. 

The spatial selection functionality is achieved by applying a  spatial constraint on the cost function of a traditional ANC system. From Eq.~(\ref{eq:e=d+wGx}), the spatial constraint \added{for a single desired source} can be expressed as 
\begin{equation}
\mathbf{H}^\text{T} \mathbf{u} = \mathbf{f},
\end{equation}
from the Frost algorithm~\citep{Frost1972}. Matrix $\mathbf{H}$ consists of the relative impulse responses (ReIRs) of the array, i.e.,
\begin{equation}
        \mathbf{H} \in \mathbb{R}^{KL \times L} = \left[ \mathbf{H}_{1} \ \mathbf{H}_{2} \ \dots \ \mathbf{H}_{K} \right]^\text{T}, 
        \label{eq:H_bold}
    \end{equation}
where $\mathbf{H}_{k}$ is the Toeplitz matrix from $\mathbf{h}_k = [{h_k}_0\ {h_k}_1\ {h_k}_2\ \dots\ {h_k}_{(L-1)} ]^\text{T}$, which is the ReIR between the $k$-th microphone and a chosen reference microphone (the one closest to the desired source). Vector $\mathbf{f} \in \mathbb{R}^{L}$ is a constraint vector that describes the frequency response of the signal from the desired direction at the error microphone. It is defined as 
    \begin{equation}
    \mathbf{f} = \mathbf{h}_K = [{h_K}_0\ {h_K}_1\ \dots\ {h_K}_{(L-1)} ]^\text{T} .
    \label{eq:f}
    \end{equation}

With ANC enabled, compared with the disturbance signal in Eq.~(\ref{eq:d_s_v}), the error signal will also contain two components,
\begin{equation}
    e(n) = e_s(n) + v_\text{ANC}(n),
    \label{eq:e=e_s+v_anc}
\end{equation}
where $e_s(n)$ is the residual desired signal and $v_\text{ANC}(n)$ is the residual noise. By choosing the appropriate reference channel for $\mathbf{H}$ and the vector $\mathbf{f}$ that corresponds to the direction of interest, the proposed system will reduce only the noise, $v(n)$, to $v_\text{ANC}(n)$. The constraint vector $\mathbf{f}$ in Eq.~(\ref{eq:f}) can enable the system to preserve the original desired sound component at the error microphone, i.e., $e_s(n) = s(n)$, instead of reconstructing it as in \citep{Serizel2010,Dalga2011,Patel2020}.

Finally, the cost function of the proposed system can be written as
    \begin{align}
    \min _{\mathbf{w}} \ E\left\{e^{2}(n)\right\} \ = \min _{\mathbf{w}} \ & E \left\{\mathbf{u}^\text{T} \Phi_{{\mathbf{x}} {\mathbf{x}}} \mathbf{u}\right\}  \notag
    \\
    \quad  = \min _{\mathbf{w}} \ & E \left\{(\tilde{\delta} + {\mathbf{G}} \mathbf{w})^\text{T} \Phi_{{\mathbf{x}} {\mathbf{x}}} (\tilde{\delta} + {\mathbf{G}} \mathbf{w})\right\}  \notag
    \\
    \text { s.t. } &\mathbf{H}^\text{T}(\tilde{\delta} + {\mathbf{G}} \mathbf{w}) = \mathbf{f}, 
    \label{eq:J_proposed_final}
    \end{align}
where $\Phi_{{\mathbf{x}} {\mathbf{x}}} \in \mathbb{R}^{KL \times KL} = E\left\{{\mathbf{x}}(n) {\mathbf{x}}^\text{T}(n)\right\}$ is the autocorrelation matrix of the tap-stacked input vector, operator $E\{\cdot\}$ denotes mathematical expectation.

\subsection{Optimal solution} \label{Section:Proposed_optimal}
The optimal solution can be found by setting the gradient of the cost function to zero and solving the Lagrange multipliers $\lambda \in \mathbb{R}^{L}$~\citep{Haykin2002}. The derivation of such an optimal solution can be found in appendix~\ref{Section:appendix_optimal}, where the solution is provided in Eq.~(\ref{eq:appendix_w_proposed_opt_full}). It can be re-written in the following form for interpretation, 
\begin{align}
    \mathbf{w}_\text{opt} = &-\Phi_{{\mathbf{r}} {\mathbf{r}}}^{-1} \phi_{{\mathbf{r}} {d}} \notag
    \\ 
    &+\Phi_{{\mathbf{r}} {\mathbf{r}}}^{-1} {\mathbf{G}}^\text{T} \mathbf{H} \left( \mathbf{H}^\text{T} {\mathbf{G}} \Phi_{{\mathbf{r}} {\mathbf{r}}}^{-1}  {\mathbf{G}}^\text{T} \mathbf{H} + \rho \textbf{I} \right)^{-1} \mathbf{f} \notag
    \\
    &- \Phi_{{\mathbf{r}} {\mathbf{r}}}^{-1} {\mathbf{G}}^\text{T} \mathbf{H} \left( \mathbf{H}^\text{T} {\mathbf{G}} \Phi_{{\mathbf{r}} {\mathbf{r}}}^{-1}  {\mathbf{G}}^\text{T} \mathbf{H} + \rho \textbf{I} \right)^{-1}  \notag
    \\
    & \qquad \qquad \qquad \qquad \left( \mathbf{H}^\text{T}\tilde{\delta} - \mathbf{H}^\text{T} {\mathbf{G}} \Phi_{{\mathbf{r}} {\mathbf{r}}}^{-1} \phi_{{\mathbf{r}} {d}}  \right) ,
    \label{eq:w_proposed_opt}
\end{align}
where 
    \begin{subequations}
        \begin{align}
           {\mathbf{r}}(n) &= {\mathbf{G}}^\text{T}{\mathbf{x}}(n),
            \label{eq:r_tilde}
        \\
             \Phi_{{\mathbf{r}} {\mathbf{r}}} &= E\left\{{\mathbf{r}}(n) {\mathbf{r}}^\text{T}(n) \right\} + \beta \textbf{I} ,
            \label{eq:Phi_rr}
        \\
             \phi_{{\mathbf{r}} {d}} &= E\{{\mathbf{r}}(n) {d}(n) \}.
            \label{eq:Phi_rd}
        \end{align}
        \label{eq:r_tilde_Phi_rr_Phi_rd}
    \end{subequations}
    
This optimal solution has three terms. The first term is the Wiener solution of a hybrid ANC system controlling all the observable sounds. The second term is due to the spatial constraint, which is similar to the beamformer solution in~\citep{Frost1972}, but the secondary path matrix ${\mathbf{G}}$ is added due to the physical constraint of the ANC system. The third term provides the coupling of the two subsystems. All three terms contribute to calculating the control filter $\mathbf{w}$ such that only the noise from the undesired directions is minimized. The residual desired signal component in the residual error signal after control is the original desired physical sound left unaltered by the proposed system. Further details about preserving or reconstructing this desired physical sound will be emphasized in Section~\ref{Section:comparison_with_existing_methods}.

Note that matrix $\mathbf{H}^\text{T} {\mathbf{G}} \Phi_{{\mathbf{r}} {\mathbf{r}}}^{-1}  {\mathbf{G}}^\text{T} \mathbf{H}$ is rank deficient due to ${\mathbf{G}}$ being rank deficient. (There are delays in the secondary paths.) Thus, a Tikhonov regularization factor $\rho$ has been applied to the diagonal elements to make it invertible~\citep{tikhonov1977}. 

Note that Eq.~(\ref{eq:Phi_rr}) has also been added with a regularization factor $\beta$ compared to Eq.~(\ref{eq:appendix_w_proposed_opt_full}), which is equivalent to adding a penalty term, the $l_2$ norm of the control filter $||\textbf{w}||_2$ in Eq.~(\ref{eq:J_proposed_final}). This is essentially the leaky version of the algorithm for a more robust ANC system~\citep{elliott1992behavior,cartes2002experimental}. The robustness of the system will be further discussed in Section~\ref{Section:robustness}.

\subsection{Adaptive solution} \label{Section:Proposed_adaptive}

\begin{figure*}[t]
        \centering
        \includegraphics[width=0.53\textwidth]{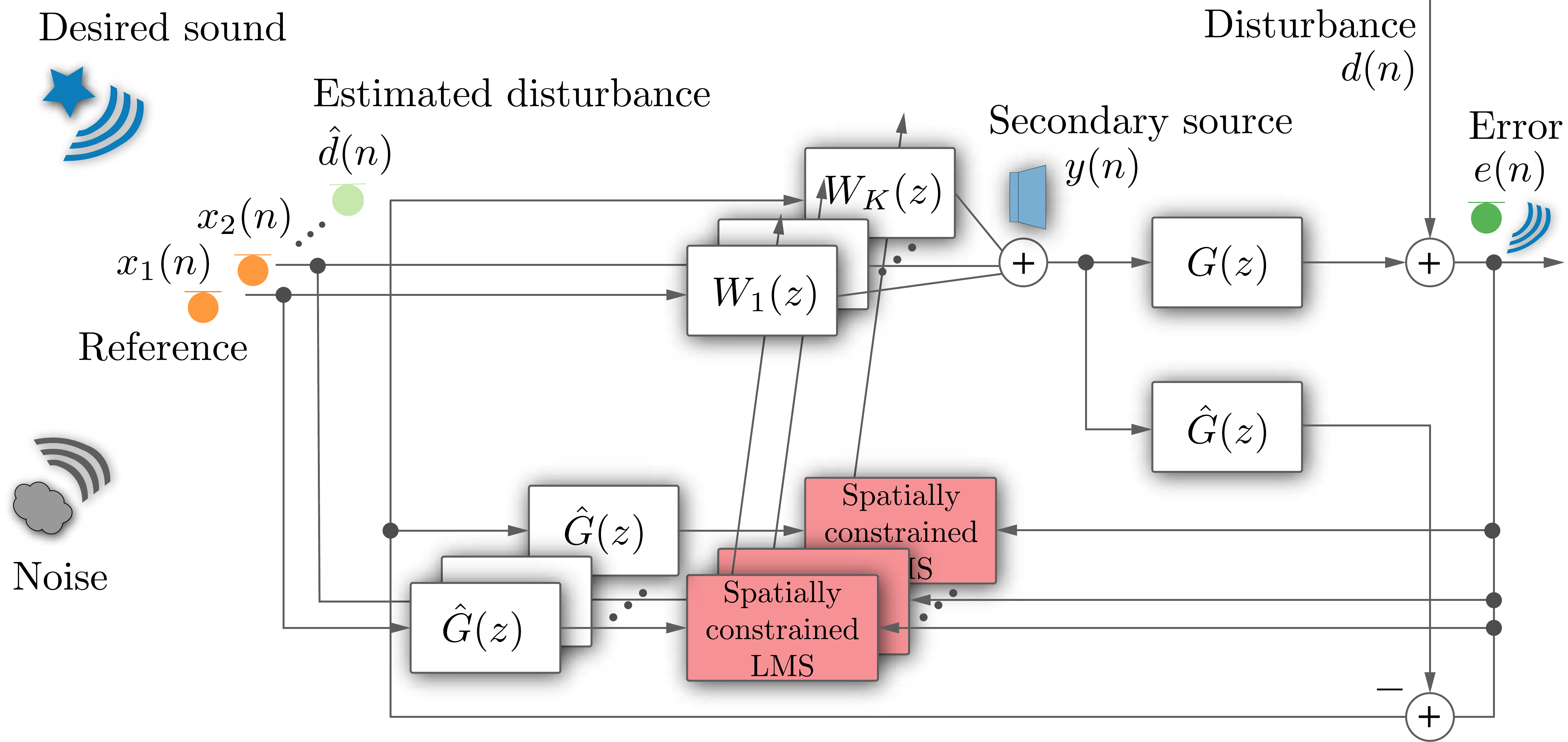}
        \caption{Block diagram of proposed spatially selective ANC system. The adaptive hybrid ANC algorithm is spatially constrained.}
        \label{fig:2}
    \end{figure*}

Commonly, adaptive algorithms are used to reduce computations and handle fast environmental changes. The derivations of the adaptive solution in the proposed method can be found in appendix~\ref{Section:appendix_adaptive}. The final solution is expressed as
    \begin{subequations}
    \begin{equation} 
    \mathbf{w}(0) = \mathbf{q}, \label{eq:w0_proposed}  
    \end{equation}
    \begin{equation} \mathbf{w}(n+1) = \mathbf{P} \left[\mathbf{w}(n) - \mu {\mathbf{G}}^\text{T} {\mathbf{x}}(n)e(n) \right] + \mathbf{q}, \label{eq:w_n+1_proposed} 
    \end{equation}
    \label{eq:w_proposed}
    \end{subequations}
where 
\begin{subequations}
\begin{align}
\mathbf{P}&=\mathbf{I}-{\mathbf{G}}^\text{T} \mathbf{H}\left(\mathbf{H}^\text{T} {\mathbf{G}} {\mathbf{G}}^\text{T} \mathbf{H}+\gamma \mathbf{I}\right)^{-1} \mathbf{H}^\text{T} {\mathbf{G}},
\label{eq:P_proposed_gamma}
\\
\mathbf{q}&={\mathbf{G}}^\text{T} \mathbf{H}\left(\mathbf{H}^\text{T} {\mathbf{G}} {\mathbf{G}}^\text{T} \mathbf{H}+\gamma \mathbf{I}\right)^{-1}\left(\mathbf{f}-\mathbf{H}^\text{T} \tilde{\delta}\right),
\label{eq:q_proposed_gamma}
\end{align}
\label{eq:P_q_proposed_gamma}
\end{subequations} 
\noindent and $\mu$ is the step-size. Fig.~\ref{fig:2} illustrates the block diagram of the proposed adaptive spatially selective ANC system.

Notice that the solution is coupled by the adaptive algorithm in the ANC subsystem and the adaptive Frost algorithm. The spatial constraint in Eq.~(\ref{eq:J_proposed_final}) resulted in $\mathbf{P}$ and $\mathbf{q}$. Without it, the solution would become $\mathbf{w}(n+1) = \mathbf{w}(n) - \mu {\mathbf{G}}^\text{T} {\mathbf{x}}(n)e(n) $, which is a traditional adaptive hybrid ANC solution minimizing the overall error signal~\citep{Elliott2000,Hansen2012}.

Notice that a regularization factor $\gamma$ in Eq.~(\ref{eq:P_q_proposed_gamma}) has been added compared to Eq.~(\ref{eq:P_q_proposed}). Similarly to the optimal solution, although the matrix $\mathbf{H}^\text{T} \mathbf{H}$ is invertible due to the use of ReIRs, that is, there is always an identity matrix in $\mathbf{H}$, matrix $\mathbf{H}^\text{T} {\mathbf{G}} {\mathbf{G}}^\text{T} \mathbf{H}$ is rank-deficient. Therefore, $\gamma$ has been added for inversion. 

\subsection{Spectral weighting} \label{Section:Spectral_weighting}
In some cases, spatial filtering alone may not be sufficient due to limitations of the array, e.g., filter length, array configuration. To deal with this, the proposed method can be further improved by applying a spectral weighting filter, i.e., 
\begin{equation}
    \mathbb{F} \in \mathbb{R}^{L} = \mathbf{S} \mathbf{h}_K
    \label{eq:F}
\end{equation}
instead of $\mathbf{f}$ in Eq.~(\ref{eq:f}). $\mathbf{S}\in \mathbb{R}^{L \times L}$ is the Toeplitz matrix of the impulse response of the spectral weighting filter to attenuate the frequency range that is not of interest. Note that $\mathbf{S}$ is a digital filter, which can be designed to be minimum-phase. A non-minimum-phase filter is not desired since it leads to delays in the error signal.

This technique is performed provided that the frequency range attenuated by the spectral filter has little overlap with that of the desired signal. Otherwise, one needs to consider the trade-off between noise reduction and signal distortion.

\subsection{Robustness} \label{Section:robustness}
Robustness is an important factor to consider. The robustness issues in ANC systems can be contributed by non-stationary inputs, low SNR, and/or secondary path changes~\citep{Elliott2000,cartes2002experimental}. Many beamforming systems are subject to numerical and/or spatial robustness issues arising from signal mismatches, which are due to mismatches between the presumed and actual relative transfer functions~\citep{vorobyov2013principles,gannot2017consolidated}. We will mainly focus on the numerical robustness issue due to signal mismatches herein. 

For the proposed joint optimization problem, the robustness of the system can be from the following aspects:
\begin{enumerate}
    \item Robustness of the ANC subsystem  \vspace{-2mm}
    \item Robustness of the beamforming subsystem  \vspace{-2mm}
    \item ANC affects the beamforming subsystem  \vspace{-2mm}
    \item Beamforming affects the ANC subsystem (due to signal mismatches)
\end{enumerate}

The former two aspects can be easily understood since each subsystem can inherently have its own robustness issues. The solutions to these issues can also be easily found in many studies. For example, the ANC subsystem can use a leaky algorithm constraining $||\textbf{w}||_2$ in the cost function~\citep{cartes2002experimental,tobias2004leaky}. As for a robust beamformer, the diagonal loading method can be used. This is achieved by constraining the white noise gain (WNG) $||\textbf{u}||_2$ in the cost function~\citep{Cox1987,li2003robust,vorobyov2013principles}. 


The latter two issues are due to the joint optimization of ANC and beamforming. The third issue arises from the secondary paths in ANC systems, particularly the delays from the acoustic propagation and the electronics. They can be coupled with the beamforming constraints resulting in an adverse effect, e.g., the rank deficiency of ${\mathbf{G}}$ as discussed previously. Regularization factors $\rho$ and $\gamma$ have been added to solve the problem~\citep{tikhonov1977,hansen2010discrete}. 

As for the fourth aspect, the beamforming constraint can be unstable due to signal mismatches and thus affect the ANC performance. This is similar to designing a robust beamformer. Constraining the WNG in the cost function can stabilize the system. The optimal solution presented previously has been applied with the regularization factor $\beta$ for robustness~\citep{Cox1987,vorobyov2013principles}.


\section{Numerical Simulation on Augmented Reality (AR) Glasses} \label{Section:simulations}
In this section, we put the proposed algorithm in a pair of open-fitting AR glasses.

\subsection{System setup} \label{Section:setup}
\begin{figure*}[t]
        \centering
        \captionsetup[subfloat]{captionskip=0pt,farskip=0pt}
        \subfloat[]{\includegraphics[height=1.8in]{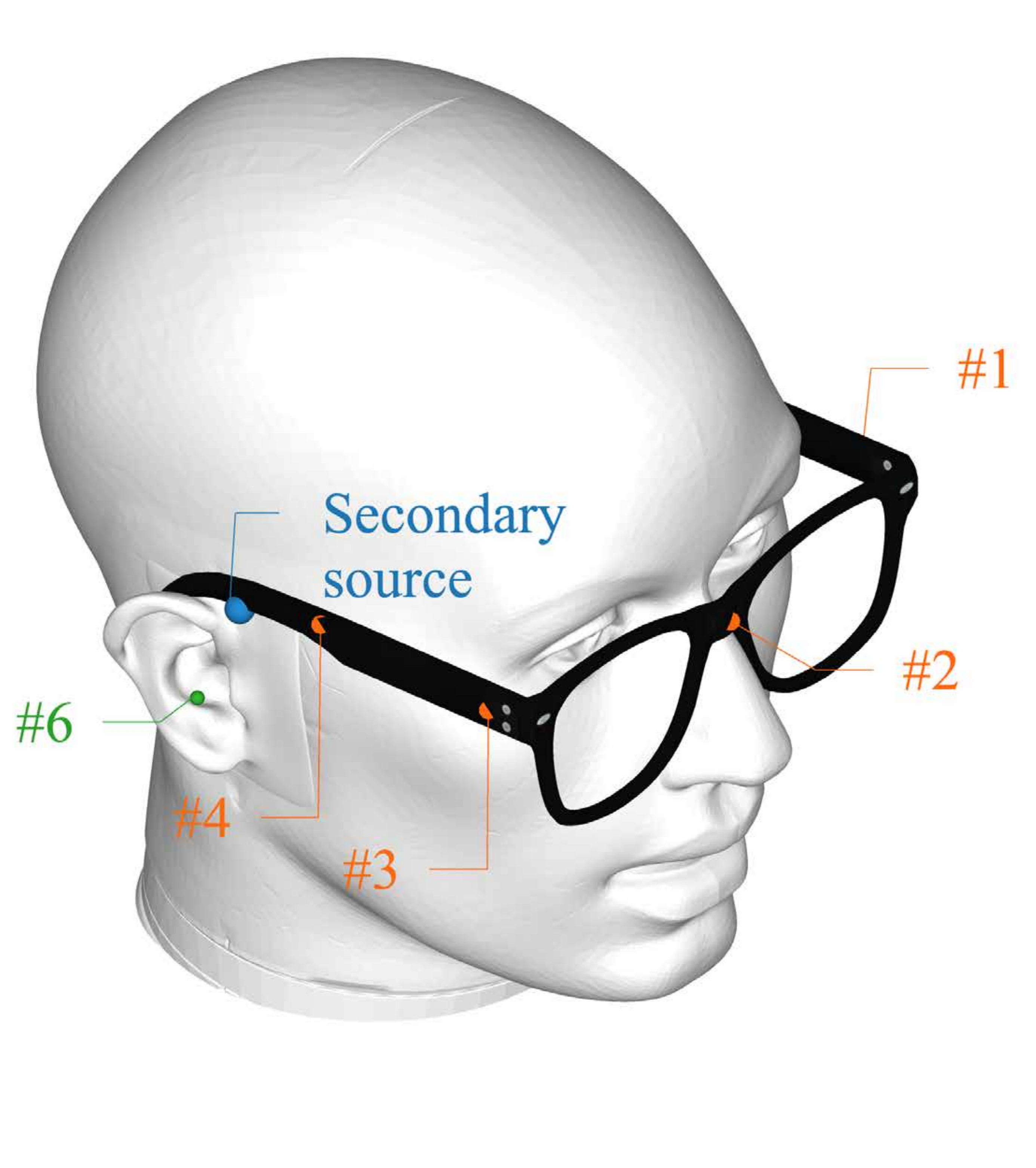}\label{fig:3a}}
        \quad \quad
        \subfloat[]{\includegraphics[height=1.8in]{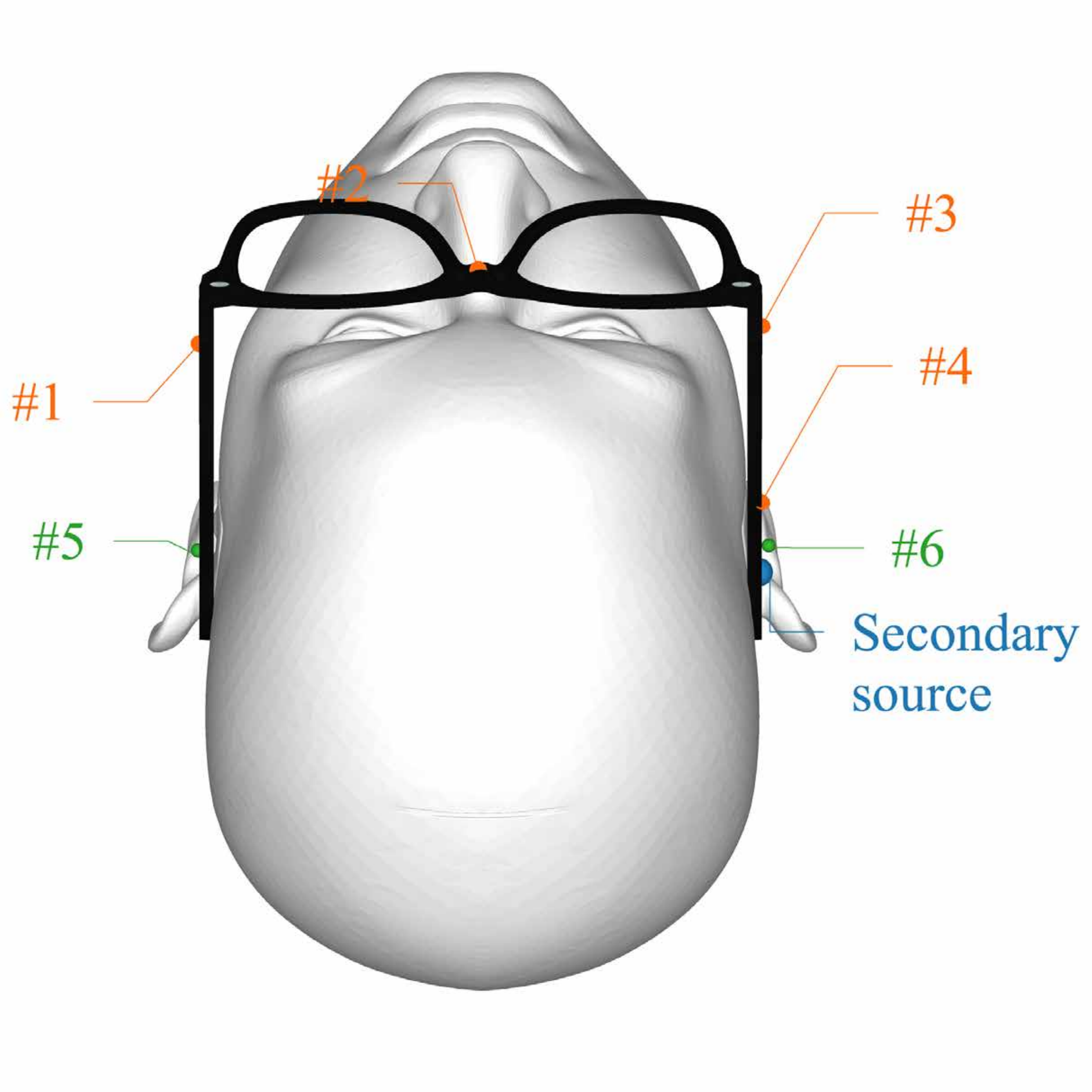}\label{fig:3b}}
        \quad \quad
        \subfloat[]{\includegraphics[height=1.8in]{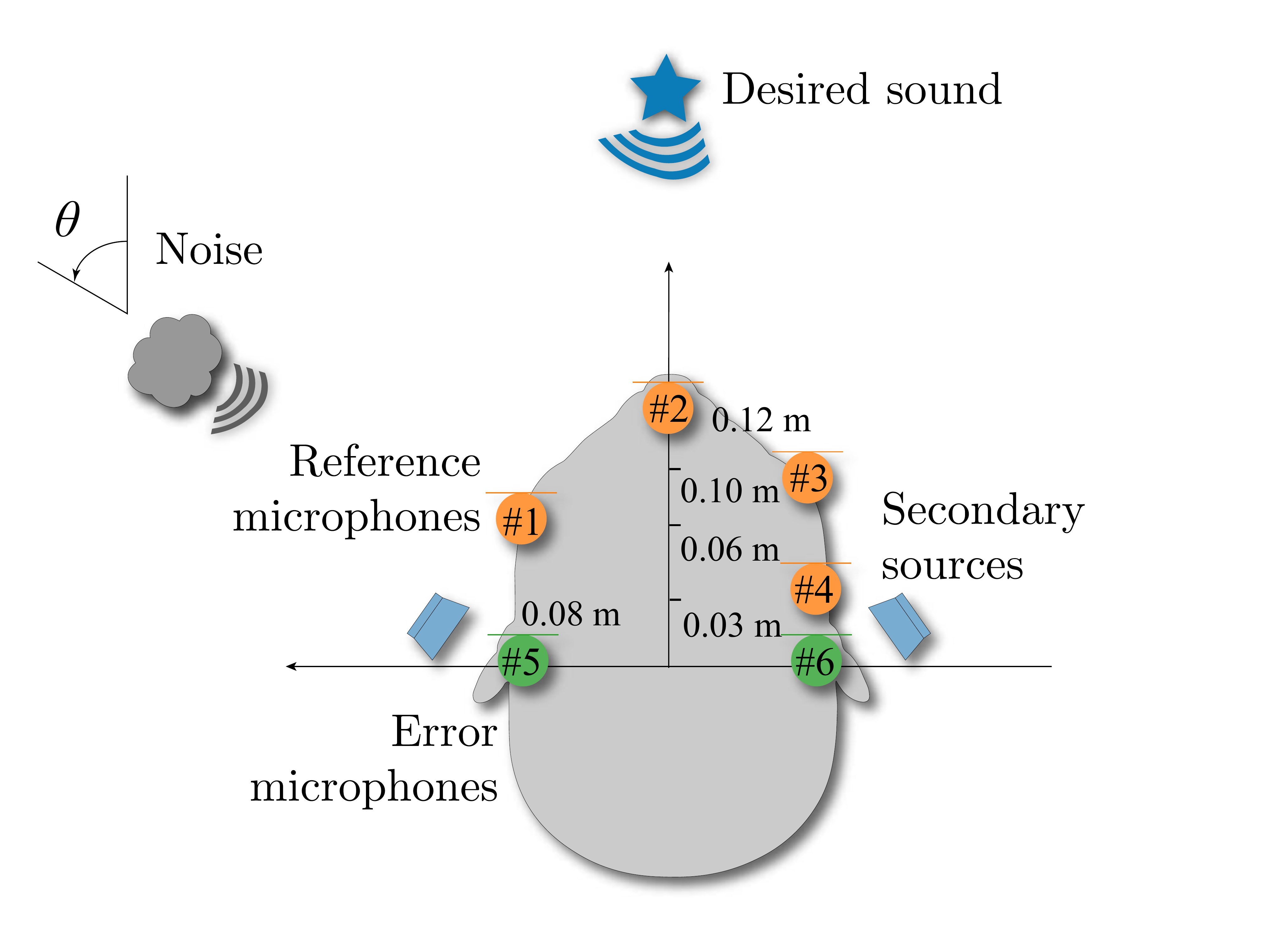}\label{fig:3c}}
        \caption{(a) The isometric view and (b) the top view of a KEMAR manikin with a pair of AR glasses with a six-microphone array. (c) Microphone setup.}
        \label{fig:3}
    \end{figure*}

A six-channel system from the EasyCom dataset~\citep{donley2021easycom} is shown in Fig.~\ref{fig:3}. The four microphones in the frame (labeled as \#1 to \#4) can be used for the reference microphones. For each ear, the corresponding binaural microphone (labeled as either \#5 or \#6) was used as the error microphone. The single-ear (right side) system is considered in this article for brevity, but it can be easily extended to a binaural case. The control performance at microphone \#6 is focused hereinafter.

\begin{figure}[t]
        \centering
        \captionsetup[subfloat]{captionskip=1pt,farskip=0pt}
        \subfloat[]{\includegraphics[width=3.35in]{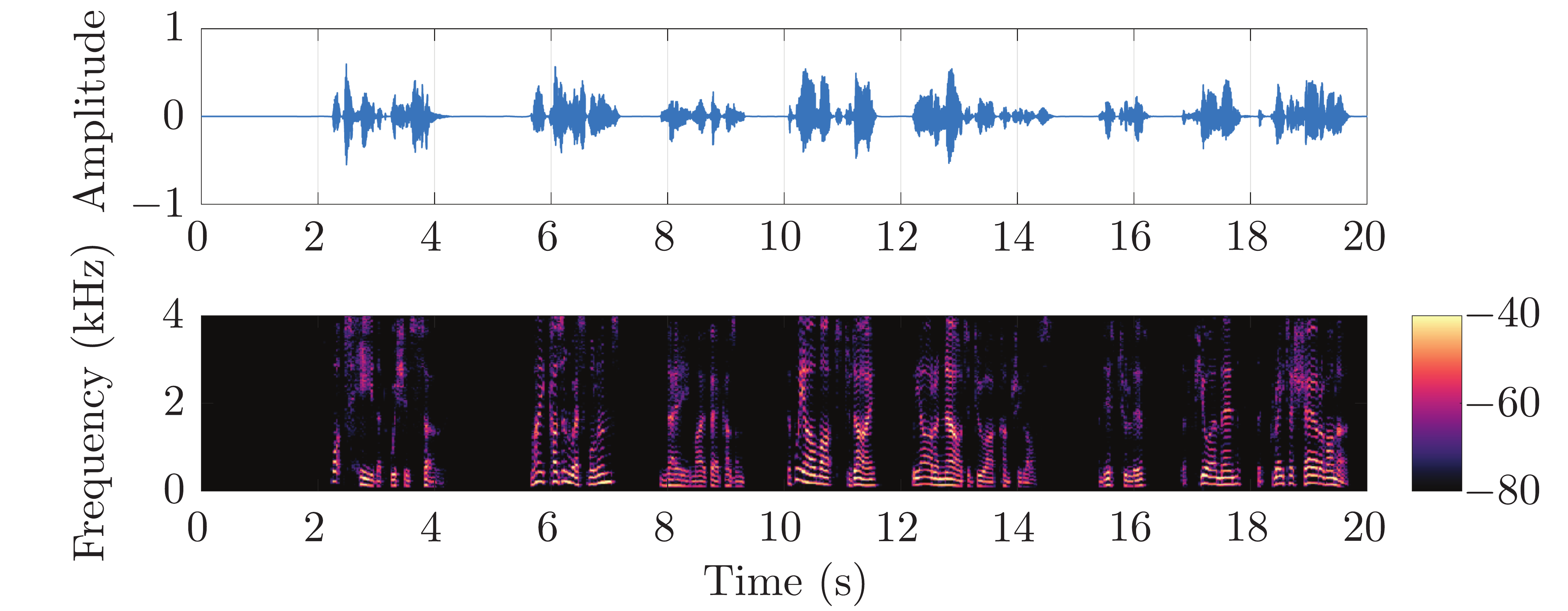}\label{fig:4a}}
        \\
        \subfloat[]{\includegraphics[width=3.35in]{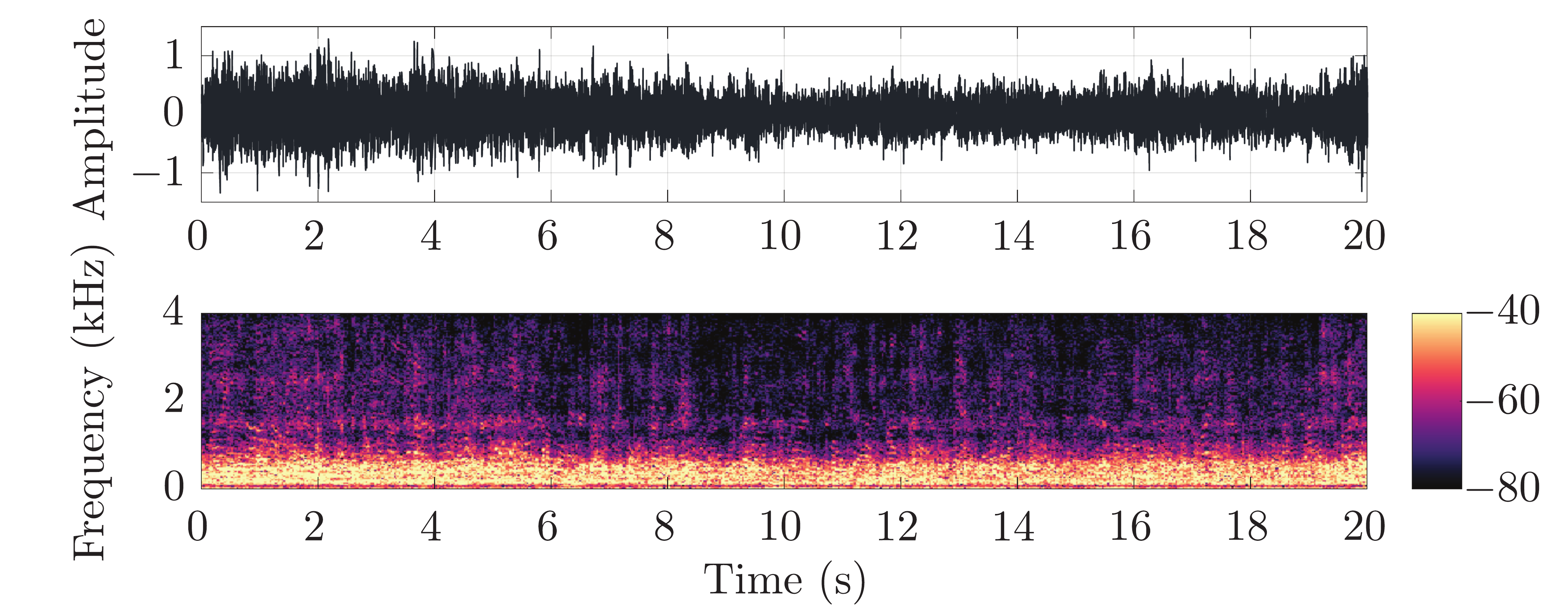}\label{fig:4b}}
        \caption{Waveforms and spectrograms of (a) the desired clean speech and (b) the noisy speech at the error microphone.}
        \label{fig:4}
    \end{figure}

It was assumed that the desired speech was at $\theta=0^\circ$ and the noise source was at $\theta=60^\circ$. The desired signal was a 20-second male speech~\citep{freesound}. The noise was a speech babble noise from the NOISEX-92 database~\citep{Varga1993}. The noise level was adjusted such that the original clean speech was not intelligible when mixed, i.e., the \textit{a priori} SNR was $-13.2$~dB. The waveforms and the spectrograms of the clean speech and the noisy speech can be seen in Figs.~\ref{fig:4a} and \ref{fig:4b}, respectively. These signals were influenced by the KEMAR manikin~\citep{Burkhard1975} and the glasses. The frequency response of the acoustic secondary path was acquired from the COMSOL Multiphysics software, where the secondary source was \added{simulated as a perfect point sound source and was} located at about 0.05~m above the error microphone \#6 as shown in Fig.~\ref{fig:3a}. In this article, we assumed \added{that the sound source was not constrained by transducer characteristics (such as excursion limit and transducer resonance), and} the response of the electrical control system and the transducers could be modeled with pure delays. The sampling rate was 48~kHz, and the filter length $L=768$. The secondary path delay had ten samples (208.3~$\upmu$s), which included both the acoustic propagation delay and the electronics delay in practice. 
%

\subsection{Performance metrics} \label{Section:simulation_perform_eval}
To quantitatively evaluate the control performance, the noise reduction (NR) level of the ANC system is defined as
\begin{equation}
    \text{NR} = 10\log_{10} \left( \frac{E\left\{v^2(n) \right\}}{E\left\{v_\text{ANC}^2(n) \right\} } \right),
    \label{eq:noise_reduction}
\end{equation}
where the NR level should be as high as possible. 

In addition, the speech distortion index (SDI)~\citep{Chen2006} is used to monitor any potential distortion in the residual speech component, and it is calculated as
\begin{equation}
    \text{SDI} = 10\log_{10} \left( \frac{E\left\{\left[s(n)-e_s(n)\right]^2 \right\}}{E\left\{s^2(n) \right\}} \right),
    \label{eq:SDI}
\end{equation}
where $e_s(n)$ should match $s(n)$ as much as possible, and thus the SDI value should be as small as possible.

Signals $e_s(n)$ and $v_\text{ANC}(n)$ can be decoupled by recording the history of the control filter $\mathbf{w}$, which is then used to re-filter either the desired speech signal or the noise. For example, one can nullify the desired speech signal while maintaining the noise to observe the NR level. Similarly, one can observe the SDI value by nullifying the noise. This method is only used to quantitatively evaluate the components in the error signal in this article. In reality, the desired signal and the noise are unknown. It is typical to estimate them at different time intervals in speech processing and then estimate the NR levels and the SDI values.

\subsection{Control performance} \label{Section:control_performance}
The step-size $\mu$ used in the simulation was a variable step-size (VSS) to ensure the system with a fast convergence speed and small misadjustment for the desired speech signal~\citep{Aboulnasr1997}. The parameters chosen for calculating the VSS were: $\mu_\text{max}=0.0001$, $\mu_\text{min}=0.000008$, $\alpha_\text{VSS}=0.99998$, $\gamma_\text{VSS}=0.00001$ and $\beta_\text{VSS}=0.99999$. The regularization factor $\gamma$ in Eq.~(\ref{eq:P_q_proposed_gamma}) was chosen to be 0.0001. A minimum-phase high-pass filter with a cut-off frequency at 140~Hz was applied as the spectral weighting filter discussed in Eq.~(\ref{eq:F}).

\begin{figure}[t]
        \centering
        \captionsetup[subfloat]{captionskip=1pt,farskip=1pt}
        \subfloat[]{\includegraphics[width=3.35in]{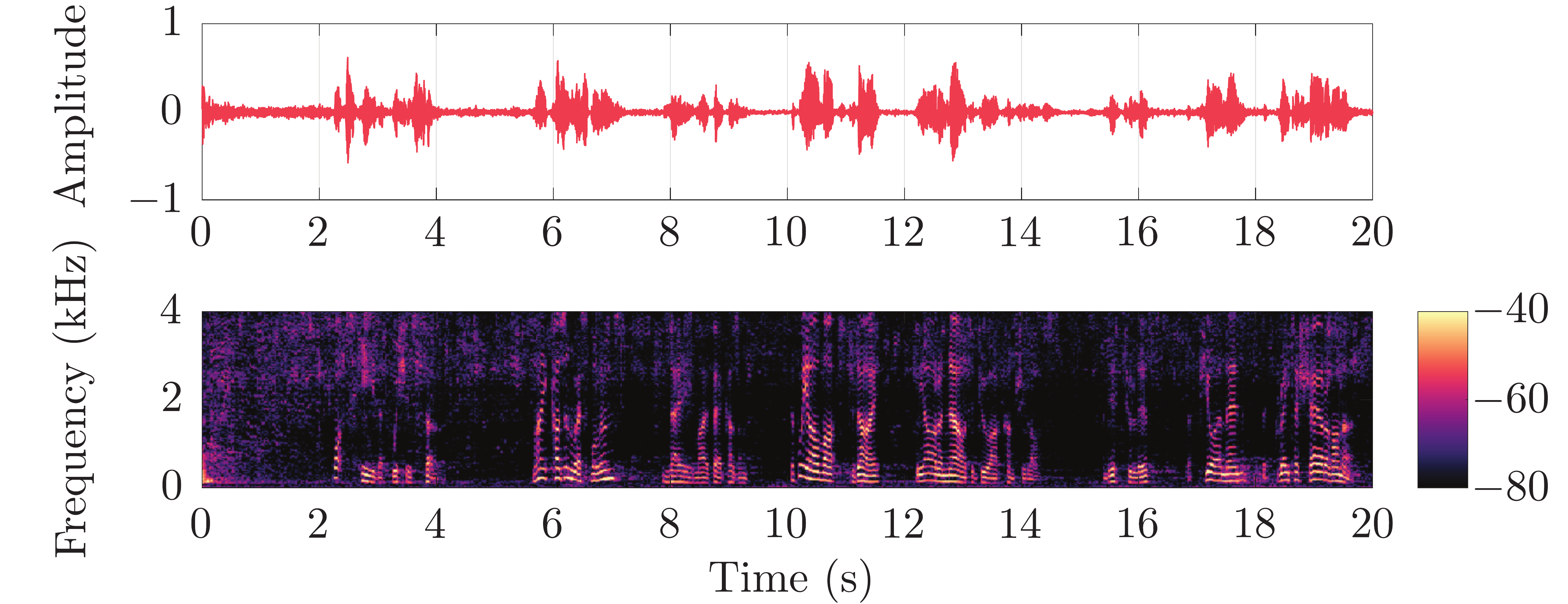}\label{fig:5a}} 
        \\
        \subfloat[]{\includegraphics[width=3.35in]{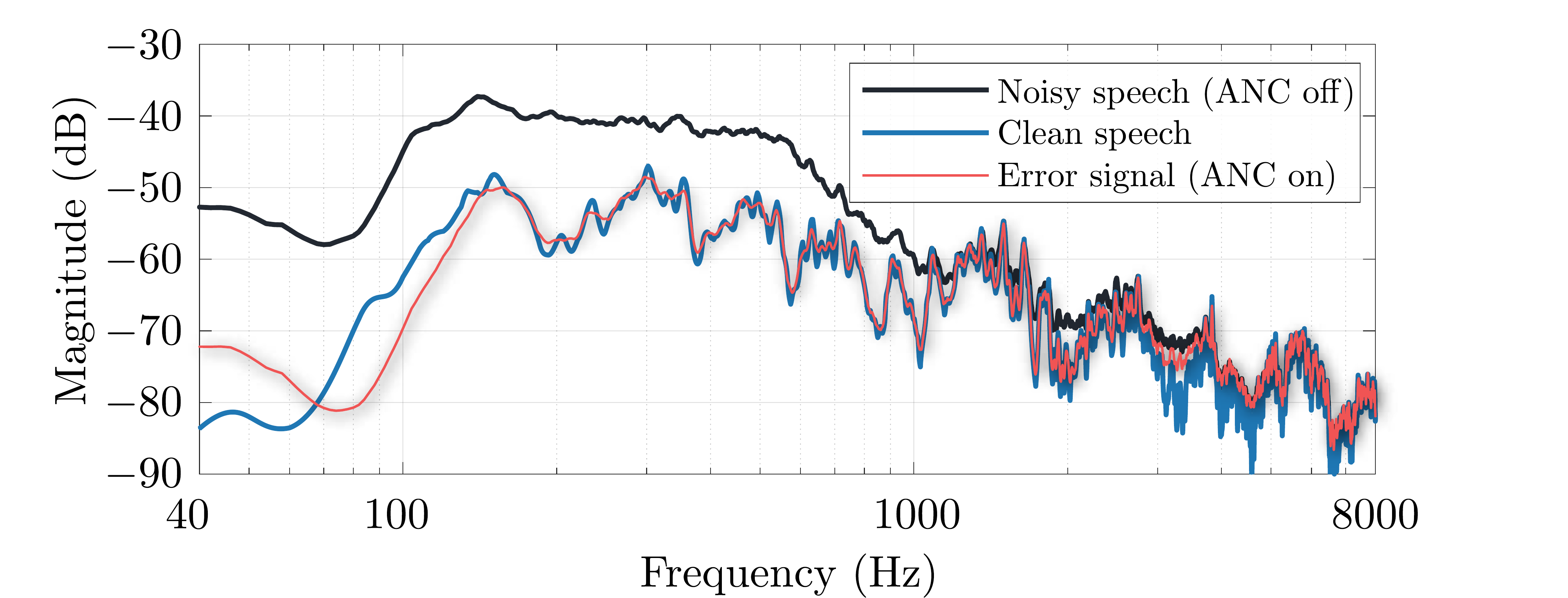}\label{fig:5b}}
        \\
        \subfloat[]{\includegraphics[width=3.35in]{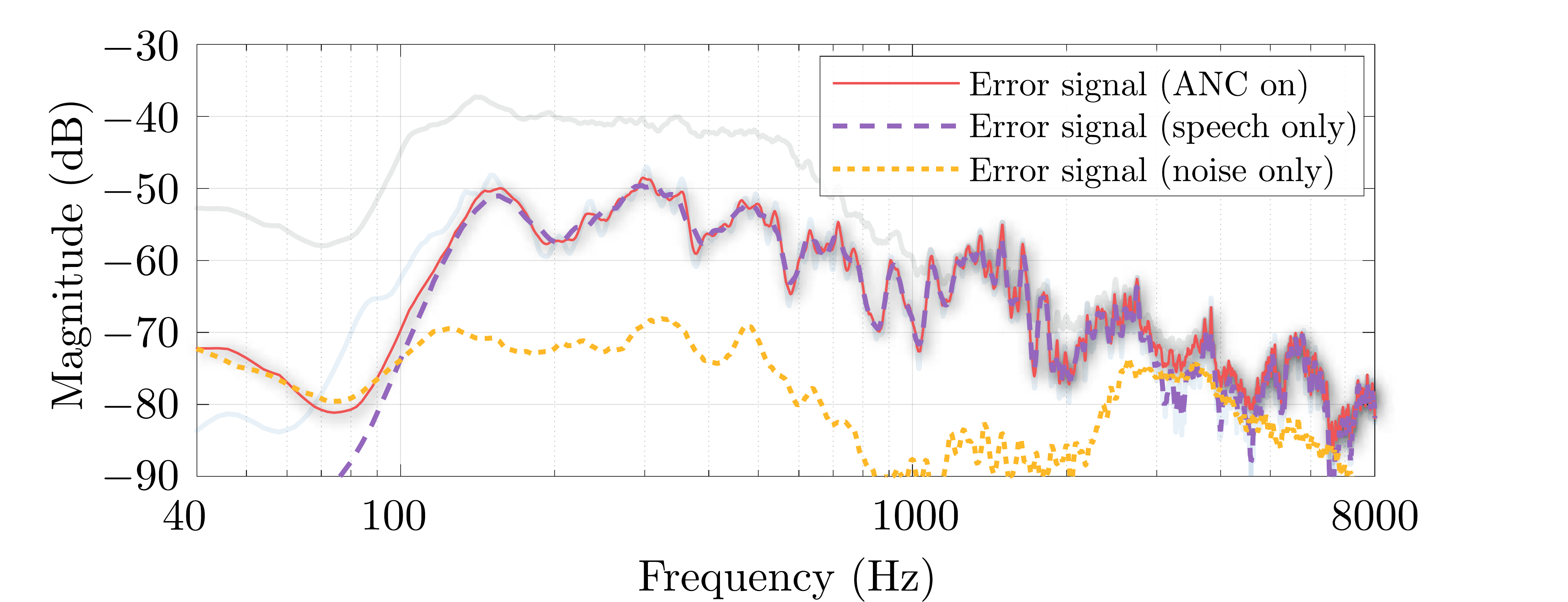}\label{fig:5c}}
        \caption{(a) Waveform and the spectrogram of the error signal with ANC enabled. (b) Spectra of the noisy speech, the clean speech, the total error signal with ANC enabled, and (c) the decoupled speech and noise components in the error signal. The spectra are from the last 10~s period.}
        \label{fig:5}
    \end{figure}

The waveform and the spectrogram of the residual error signal are shown in Fig.~\ref{fig:5a}. After ANC was enabled using the proposed method, the noise component was considerably attenuated within the first two seconds, leaving only the desired speech signal in good agreement with the original clean speech in Fig.~\ref{fig:4a} overall. 

The spectra of the noisy speech, the clean speech and the total residual error signal with ANC enabled are shown in Fig.~\ref{fig:5b}. Although the spectrum of the error signal followed the majority of the clean speech, there was still some minor residual noise below 100~Hz. As shown in Fig.~\ref{fig:5c}, decoupling the speech and the noise components as described in Section~\ref{Section:simulation_perform_eval} confirms that the system was mainly bound by the ANC subsystem. From the last 10~s of the signals, the SNR has been improved from $-13.9$~dB to 15.2~dB in total, which enhanced the unintelligible speech significantly. The NR level was 29.1~dB, and the SDI value was shown to be --25.1~dB above 100~Hz. The SDI was low enough for the listener not to notice any undesired distortion. 

\subsection{Spectral weighting} \label{Section:Spectral_wt}
\begin{figure}[t]
        \centering
        \captionsetup[subfloat]{captionskip=0pt,farskip=0pt}
        \includegraphics[width=3.4in]{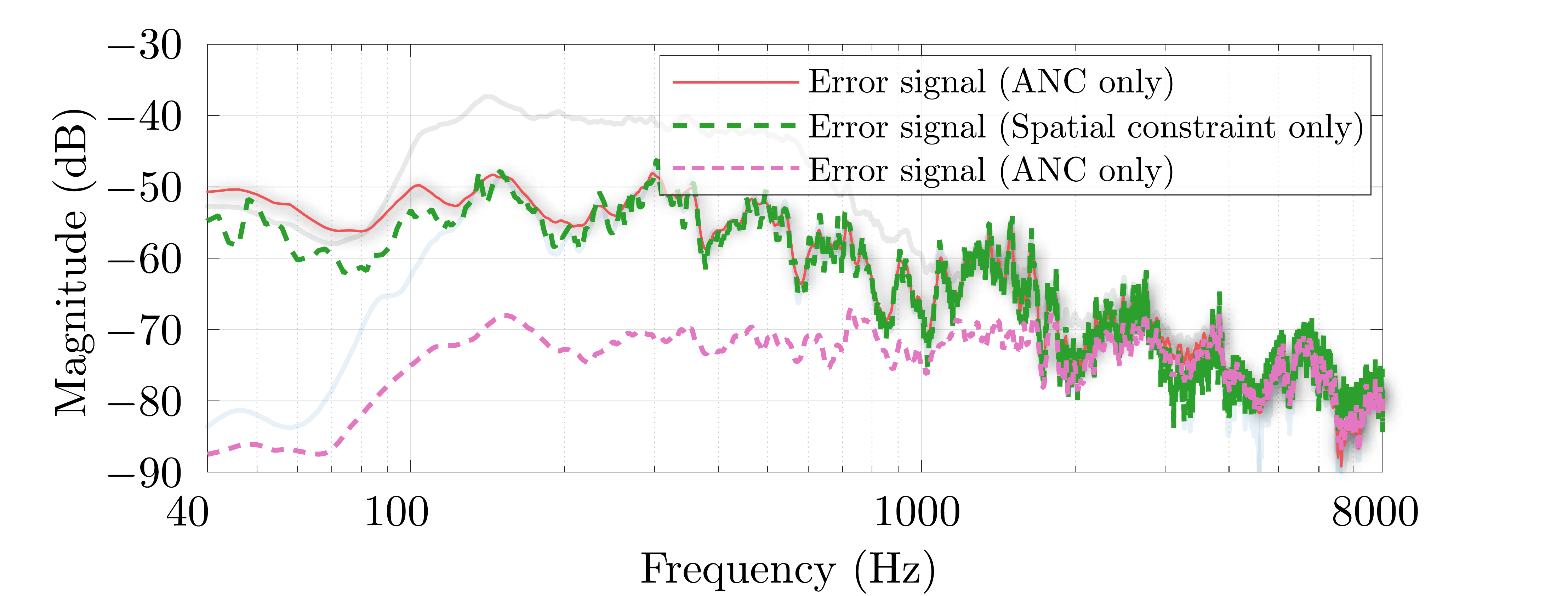}
        \caption{Spectra of the error signal from the proposed method, the error signal from the spatial constraint only and from the ANC system only after control \textit{without} using the minimum-phase high-pass filter with a cut-off frequency at 140~Hz.}
        \label{fig:6}
    \end{figure}

As discussed in Section~\ref{Section:Spectral_weighting}, the performance of the spatial constraint may be sub-optimal due to either an insufficient number of channels or limited filter lengths. Fig.~\ref{fig:6} shows the control performance without the spectral weighting filter. The error signal cannot be controlled adequately below 100~Hz. The reason can be found by performing the hybrid ANC and the Frost algorithms separately with the same six microphones. Although the ANC subsystem can reduce the noise across the spectrum using the hybrid control, the spatial constraint cannot below 100~Hz due to the limited length of the filters. Therefore, it is necessary to use the spectral weighting method. 

\subsection{Robustness} \label{Section:robustness_simulation}
    \begin{figure}[t]
        \centering
        \captionsetup[subfloat]{captionskip=2pt,farskip=6pt}
        \subfloat[]{\includegraphics[width=3.2in]{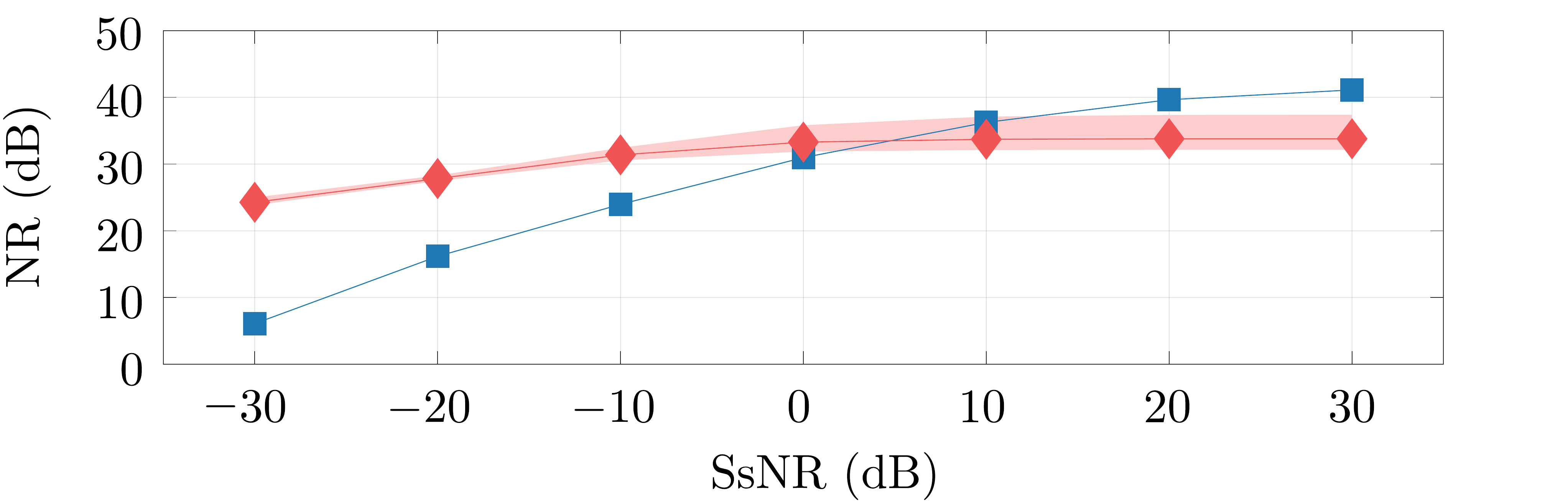}\label{fig:7a}}
        \\
        \subfloat[]{\includegraphics[width=3.2in]{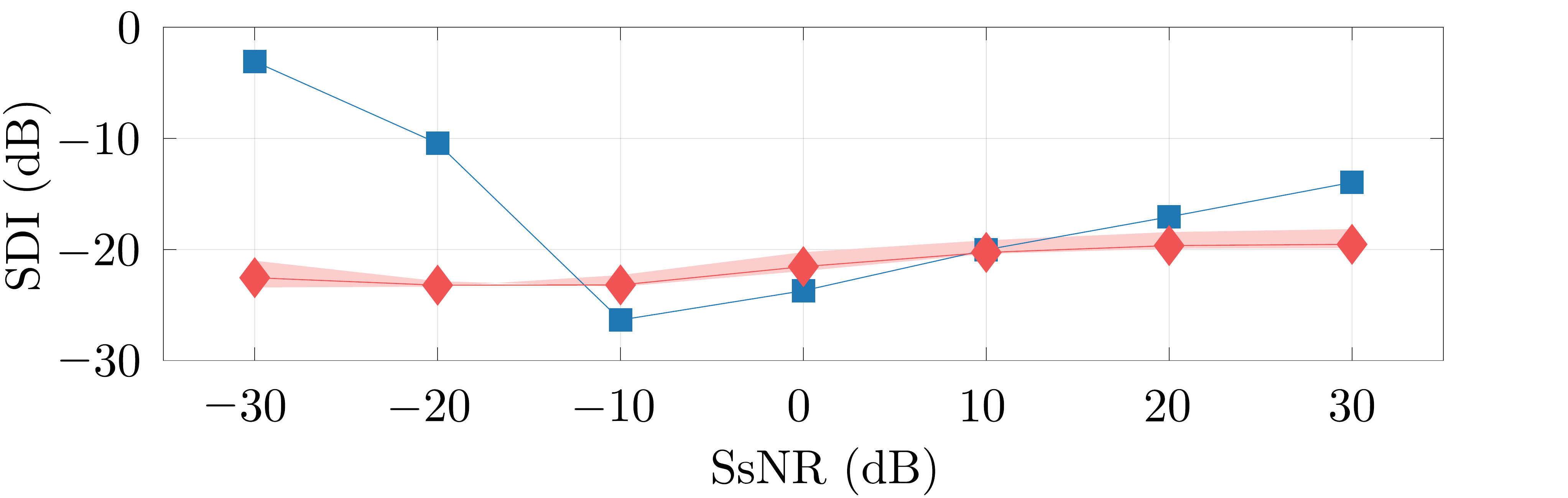}\label{fig:7b}}
        \caption{(a) The NR level and (b) the SDI value of the error signal with respect to different SsNRs. The blue square marks are the results by choosing $\beta=\rho=10\sigma_\text{n}^2$. The red diamond marks are the results by choosing $\beta=\lambda_\text{max}^1 / 10000$ and $\rho=\lambda_\text{max}^2 / 10000$. The shaded areas represent the results with the ratio from 5000 to 50000. }
        \label{fig:7}
    \end{figure}
    
This part follows the system robustness discussions in Section~\ref{Section:robustness}. We mainly examine how sensor noise (signal mismatches) from the beamforming constraint affects the ANC system and the control performance. To examine the robustness of the system, sensor noise was added to some channels, e.g., microphone \#1, \#3, \#4 and \#5. It is common to model the sensor noise as Gaussian white noise~\citep{VanTrees2002} with the power of $\sigma_\text{n}^2$. The signal-to-sensor-noise-ratio (SsNR) at microphone \#5 was used to represent different levels of sensor noise. The optimal solution in Eq.~(\ref{eq:w_proposed_opt}) was used to calculate the ANC control filter. Here, we show the importance of $\beta$ and $\rho$ for the robustness of the system.

Beamforming problems typically follow the rule of $10\sigma_\text{n}^2$ to choose the regularization factor~\citep{li2003robust,shahbazpanahi2003robust,vorobyov2003robust}. Fig.~\ref{fig:7} (blue square mark) shows the NR and SDI results of the proposed ANC system using $\beta=\rho=10\sigma_\text{n}^2$ under different SsNRs. The problem with this method is that, for example when the sensor noise is small, e.g., SsNR = 30~dB, $\beta$ and $\rho$ are too small to have a regularization effect. Although the noise can be significantly reduced (NR = 41.1~dB), the desired speech is also highly distorted (SDI = --14.0~dB). On the other hand, when the sensor noise is great, e.g., SsNR = --30~dB, $\beta$ and $\rho$ are very large and can over-regulate the system. The system can neither control noise sufficiently (NR = 6.1~dB), nor retain the original desired speech (SDI = --3.1~dB).

We show that $\beta$ and $\rho$ can be chosen depending on the largest eigenvalue of the matrix to be inverted. Assume the largest eigenvalues of $ E\left\{{\mathbf{r}}(n) {\mathbf{r}}^\text{T}(n) \right\} $ and $\mathbf{H}^\text{T} {\mathbf{G}} \Phi_{{\mathbf{r}} {\mathbf{r}}}^{-1}  {\mathbf{G}}^\text{T} \mathbf{H}$ are $\lambda_\text{max}^1$ and $\lambda_\text{max}^2$, respectively. Typically, the ratios between $\lambda_\text{max}^1$ and $\beta$ and between $\lambda_\text{max}^2$ and $\rho$ depend on different systems. Here, we found that the ratio between 5000 and 50000 had good results. The performance under this range is shown in Fig.~\ref{fig:7} with shaded areas, and the result with $\beta=\lambda_\text{max}^1 / 10000$ and $\rho=\lambda_\text{max}^2 / 10000$ is depicted as an example (red diamond mark). It is obvious the system performance has been maintained well across different levels of sensor noise. Particularly when the input signals have been disastrously perturbed, i.e., SsNR = --30~dB, the system could still exhibit good behavior. The NR level had 24.3~dB and the SDI was maintained at --22.5~dB. Thus, by choosing the regularization factors based on the largest eigenvalue instead of the sensor noise power, the proposed system can achieve a good result even with extreme cases of sensor noise.

\subsection{Directivity} \label{Section:directivity}
    \begin{figure}[t]
        \centering
        \captionsetup[subfloat]{captionskip=0pt,farskip=0pt}
        \subfloat[]{\includegraphics[width=3.22in]{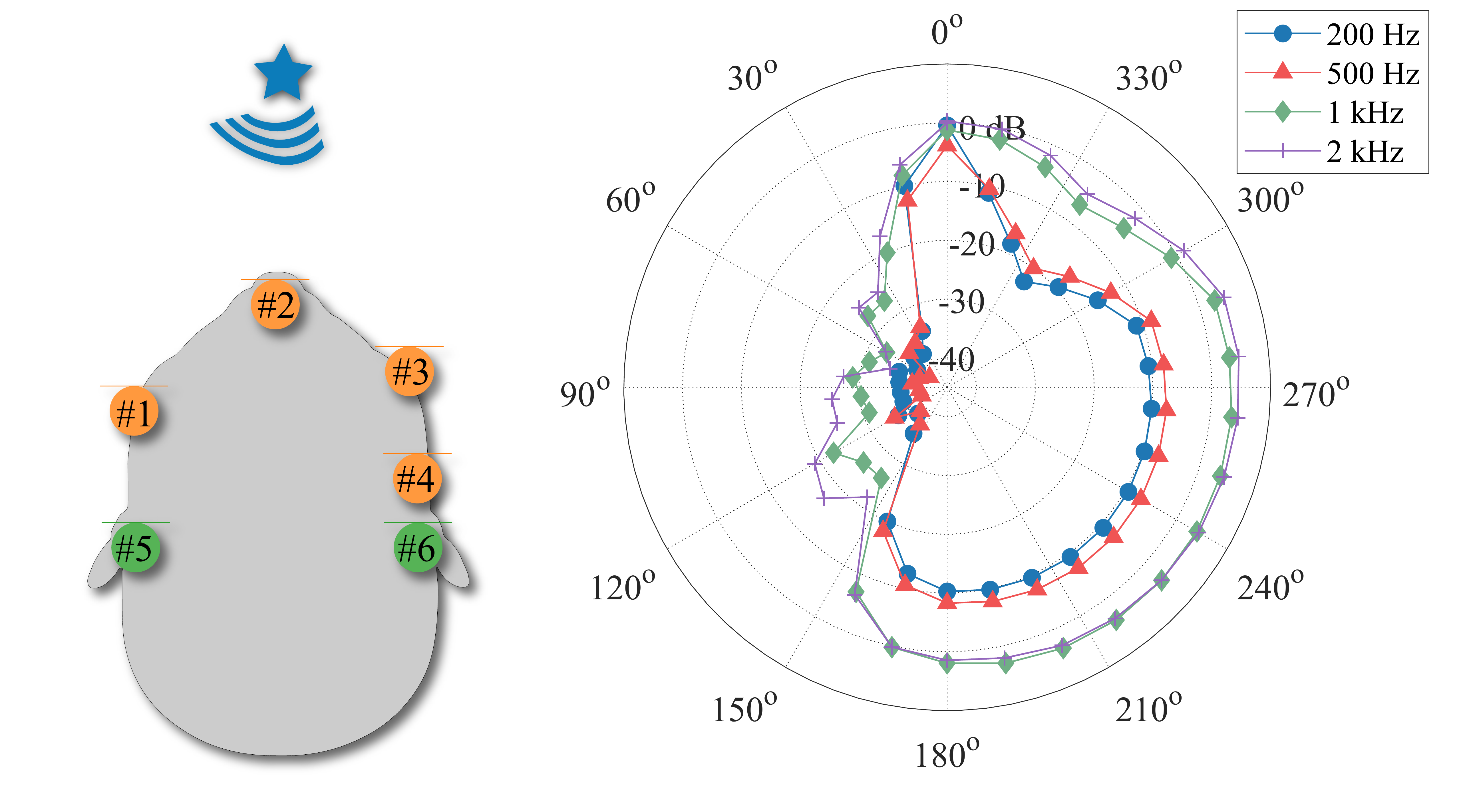}\label{fig:8a}}
        \\
        \subfloat[]{\includegraphics[width=3.22in]{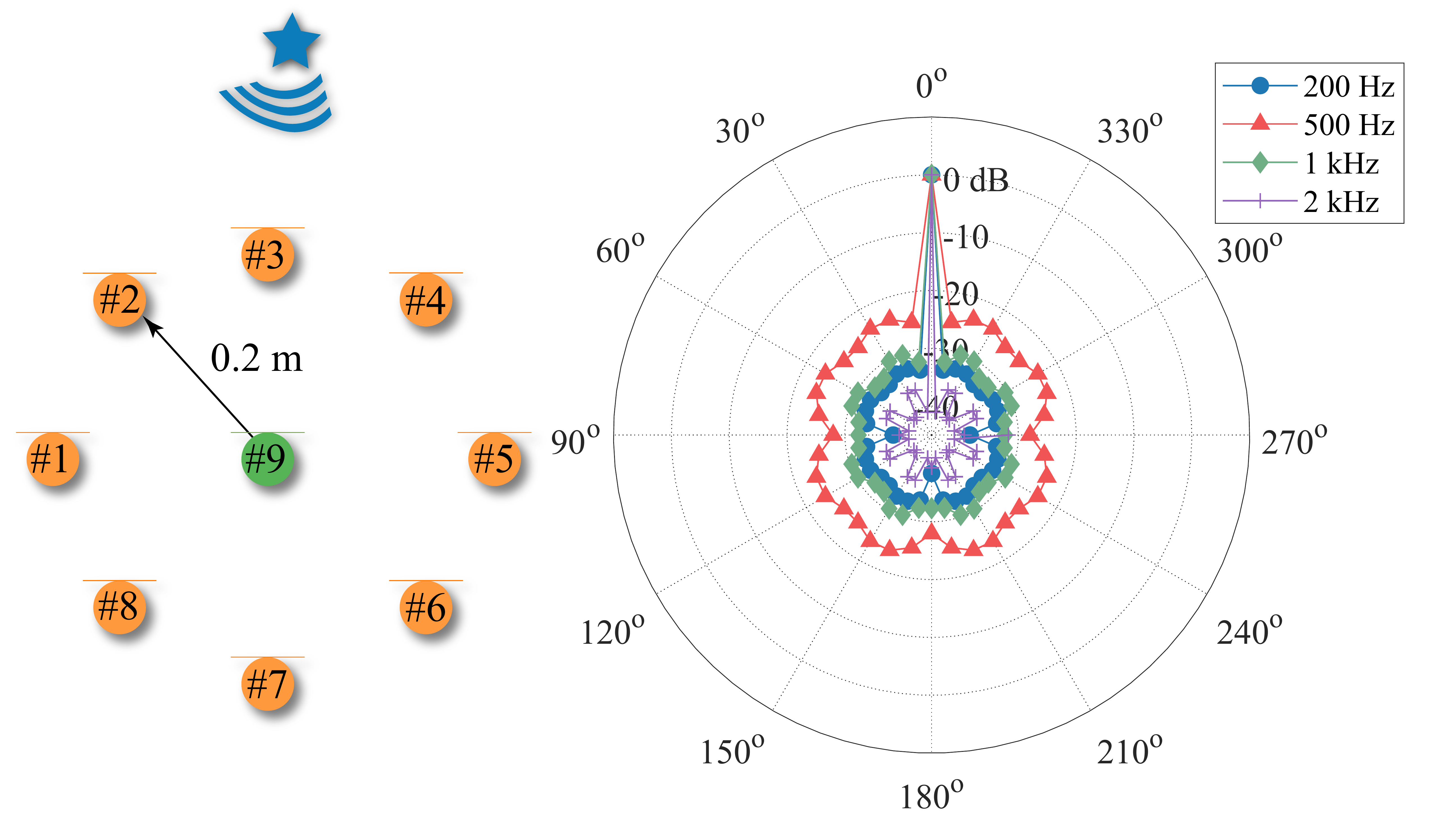}\label{fig:8b}}
        \caption{Directivity plot of the residual noise at the error microphone \#6 for a single noise source by (a) the demonstrated AR glasses configuration and (b) a system with eight reference microphones in a circular formation with an error microphone \#9 at the center for the pink noise.}
        \label{fig:8}
    \end{figure}
The direction-dependent NR performance of the demonstrated AR glasses ANC system is shown in Fig.~\ref{fig:8a} for different frequency bands. In this experiment, the desired sound source is fixed at the $0$ degree angle, and a pink noise source is placed at various angles in the horizontal plane. \added{The optimal solutions were calculated.}

At the desired direction $\theta = 0^\circ$, all signals were maintained including the noise. The NR performance was the best at $\theta \in \left(30^{\circ},\ 150^{\circ} \right)$. The noise could be reduced by at least 20~dB. Low frequencies, e.g., below 500~Hz, had more than 30~dB reduction. The system had an unsatisfactory performance at $\theta \in \left(150^{\circ},\ 300^{\circ} \right)$. This was due to the ANC capability of the specific microphone array configuration. In these directions, the noise reached the error microphone first and then the reference microphones. Thus, the system causality was violated and only the feedback subsystem was in operation, which was only effective below 500~Hz with about 10~dB reduction. High frequencies were slightly increased due to the waterbed effect~\citep{skogestadPostlethwaite2005}. 

One of the potential solutions is to decrease the delay in the secondary path, which mainly depends on the electronic components in the device. Another way is to adjust the array configuration. As shown in Fig.~\ref{fig:8b}, eight reference microphones are in a circular formation with an error microphone at the center. Although ideal, it demonstrated that it is possible to cancel noise in every direction except for the desired direction if the causality of the ANC subsystem can be maintained. Depending on the specific application, the array design will change, and the directivity pattern will change accordingly. %


\section{Comparison with existing methods}  \label{Section:comparison_with_existing_methods}
One unique feature of the proposed method is to truly preserve the original desired physical sound rather than reconstruct it. The results from the proposed method will be compared with the ones from the previous works~\citep{Serizel2010,Dalga2011,Patel2020}.

We would like to highlight the fact that the previous works had been developed for different systems with different ANC and beamforming algorithms. For ANC, \citet{Serizel2010} and \citet{Patel2020} used the feedforward configuration, while \citet{Dalga2011} used the hybrid control. For the beamforming, \citet{Serizel2010} and \citet{Dalga2011} used the MWF, while \citet{Patel2020} used the superdirective beamformer. For consistency in this article, we evaluated all the methods for the aforementioned open-fitting AR glasses with the fixed six-array microphones. The monaural setup at microphone~\#6 is still generally considered for simplicity of discussion. For a fair comparison, the hybrid ANC control and the Frost algorithm were used for all the cases. \added{The optimal solutions were also computed for most cases, except Section~\ref{Section:multiple_noises}.} \textit{The main goal for the comparison was to see how the desired signal was obtained (e.g., reconstructed or preserved) in these systems, and their implications. } 

The three configurations were configured as follows:
\begin{enumerate} 
    \item For~\citep{Serizel2010,Dalga2011}, the ANC and beamformer modules can be \textit{partially coupled}. Microphones \#1 to \#6 were used for ANC, and two microphones (\#2 and \#4) were also used as the beamforming array. The extracted signal from the beamformer had 1~ms delay, and the gain was 0~dB as provided by~\citet{Serizel2010}. The extracted signal was added to the error signal [see Fig.~5 in \citep{Serizel2010}].
    \vspace{-2.5mm}
    \item For the \textit{decoupled} configuration by~\citet{Patel2020}, microphones \#1, \#3 were used as the reference microphones for the ANC subsystem, and microphones \#2 and \#4 were used as the beamforming array. The other error microphone~\#5 was not available to control~\#6. Note that such a decoupled configuration requires dedicated reference microphones with error microphones for ANC (microphone \#1 with \#5 for the left channel ANC, \#3 with \#6 for the right). Thus, only microphones \#2 and \#4 are left for the beamformer. The extracted signal from the beamformer had 5~ms delay. The extracted signal was added to the secondary source [see Fig.~3 in \citep{Patel2020}].    
    \vspace{-2.5mm}
    \item The proposed method, as described in the previous section. 
\end{enumerate}

The aforementioned three aspects - control effort of the secondary source, control performance for multiple noise sources and binaural localization cues and latency of the desired sound - are discussed below.

\subsection{Control effort}
\deleted{Typically $||\textbf{w}||_2$ in frequency domain is used to evaluate the system control effort~(Elliott \textit{et al}., 1992, 2012). Here, we suggest the secondary source signal $y(n)=\mathbf{w}^\text{T}{\mathbf{G}}^\text{T}{\mathbf{x}}(n)$ in time domain to be more straightforward.}

\begin{figure}[t]
    \centering
    \captionsetup[subfloat]{captionskip=1pt,farskip=2pt}
    \subfloat[]{\includegraphics[width=3.5in]{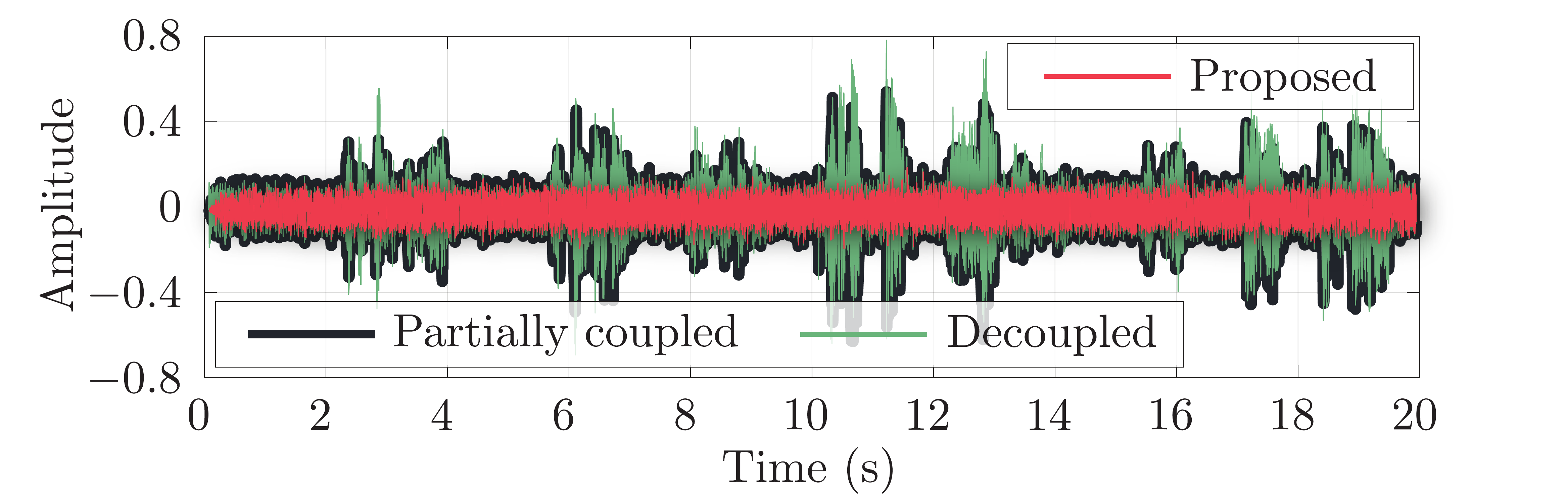} \label{fig:9a}}
    \\
    \subfloat[]{\includegraphics[width=3.5in]{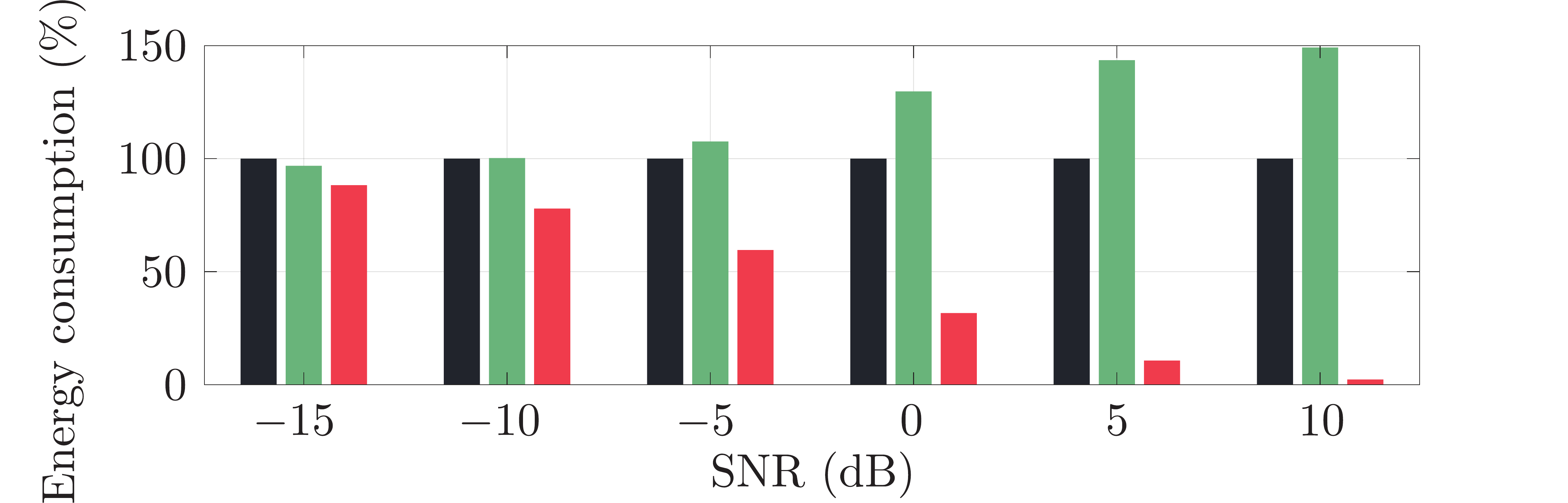} \label{fig:9b}}
    \\
    \subfloat[]{\includegraphics[width=3.5in]{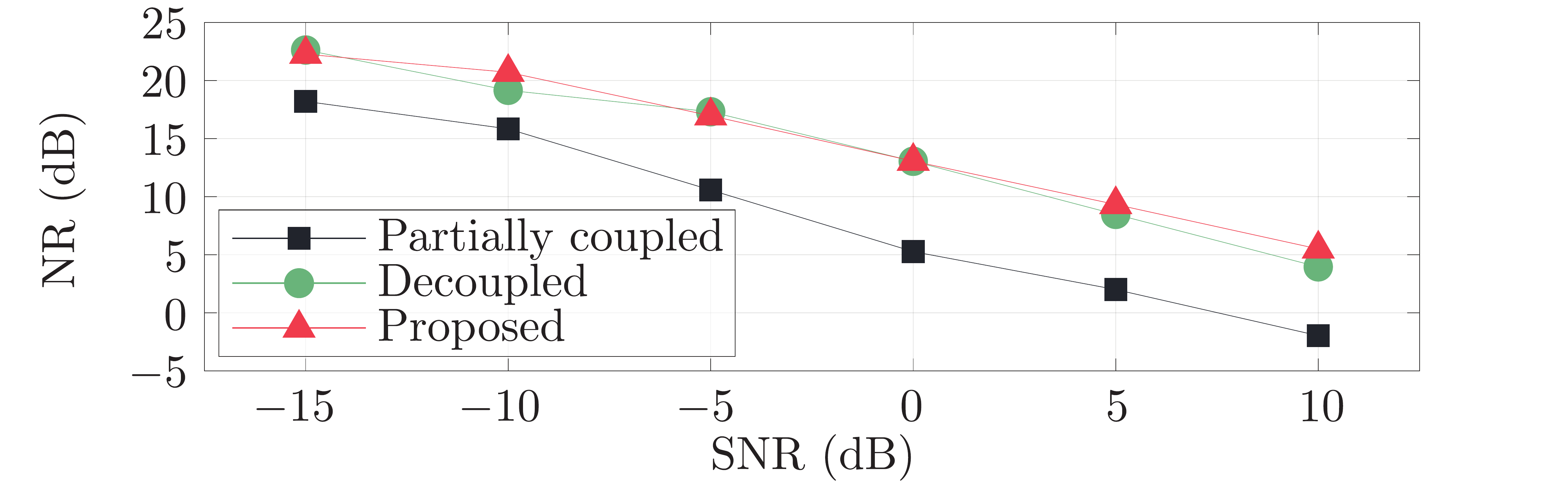} \label{fig:9c}}
    \\
    \caption{(a) Secondary source signals from the three configurations when the \textit{a priori} SNR = 0~dB. (b) Relative secondary source energy consumption, and (c) the NR levels for different \textit{a priori} SNRs.}
     \label{fig:9}
 \end{figure}

Fig.~\ref{fig:9a} shows the secondary source signal $y(n)$ in the three systems when the desired speech was at $0^\circ$, the pink noise was at $60^\circ$, and the \textit{a priori} SNR was 0~dB. It is clear that the secondary source signals in \replaced{configurations 1 and 2}{the partially coupled and the decoupled configurations} contained the desired speech. These systems needed to cancel both the desired speech and noise first and then reconstruct and reproduce the desired speech. It is even more so for \replaced{Configuration 2}{the decoupled configuration} since the desired speech needs to be directly injected into the secondary source signals [see Fig.~3 in \citep{Patel2020}]. On the contrary, the one from the proposed method was only for the noise, indicating it controlled the noise only and thus preserving the original desired speech.

\replaced{Another way for confirmation is to compare the secondary source energy for various \textit{a priori} SNRs. Using configuration~1 as the reference, the secondary source energy consumption in percentage was calculated as}
{Another way for confirmation is to compare the control effort for various \textit{a priori} SNRs. We used the secondary source energy $E\{y^2(n)\}$ instead of $||\textbf{w}||_2^2$ since the former provides a clearer physical representation~\citep[p.~147]{Elliott2000}. The relative energy consumption $\mathscr{E}_y$ in percentage was calculated as}
\begin{equation}
    \mathscr{E}_y = \frac{E\{y^2(n)\}}{E\{y_\text{ref}^2(n)\}} \times 100\% ,
    \label{eq:P_y}
\end{equation}
where $E\{y_\text{ref}^2(n)\}$ is \replaced{the total energy consumed}{the mean-square of the secondary source signal} in \replaced{configuration~1}{\citep{Serizel2010, Dalga2011}} \added{for a time period (e.g., 20 seconds) as the reference}.

The energy of the secondary source and the corresponding NR levels are shown in Figs.~\ref{fig:9b} and \ref{fig:9c}, respectively. When the noise level was high, e.g., SNR = $-15$~dB, the energy difference was small since most energy was devoted to controlling the noise for all cases. The NR levels were also similar. However, as the desired speech became more and more prominent, the difference became more and more apparent. For example, when SNR = 10~dB, the environment is relatively quiet. \replaced{Configurations 1 and 2}{The other two configurations} still used a vast amount of energy. Particularly, \replaced{configuration 2}{the decoupled configuration} used 50\% more energy than \replaced{configuration 1}{the partially coupled} due to better control performance. (The reason for this will be discussed in the next subsection.) These systems see the desired speech as ``noise'' too and attempt to cancel it first, and then try to reconstruct the desired speech once again.

On the other hand, the proposed algorithm barely needed to change the original desired signal as illustrated in the time domain in Fig.~\ref{fig:9a}. It required only 2\% energy while yet achieving a better NR level than \replaced{configuration~1}{the partially coupled configuration}. These results further confirm that the proposed system only reduces the noise and truly \textit{preserves} the desired sound.

\subsection{Noise attenuation for multiple noise sources} \label{Section:multiple_noises}
    \begin{figure*}[t]
        \centering
        \captionsetup[subfloat]{captionskip=1pt,farskip=0pt}
        \subfloat[Overall]{\includegraphics[width=1.85in]{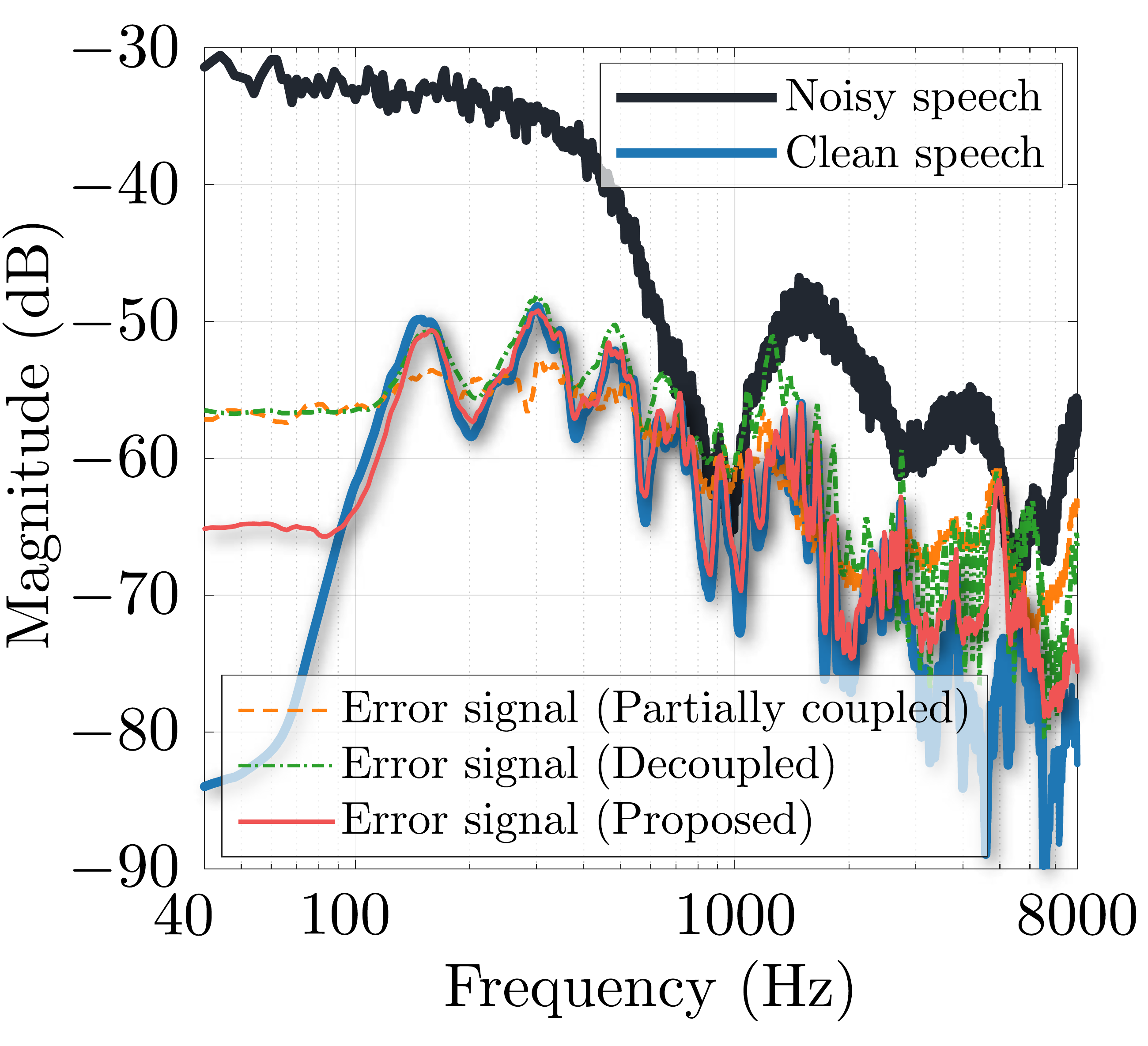}\label{fig:10a}}
        \qquad
        \subfloat[Noise component]{\includegraphics[width=1.85in]{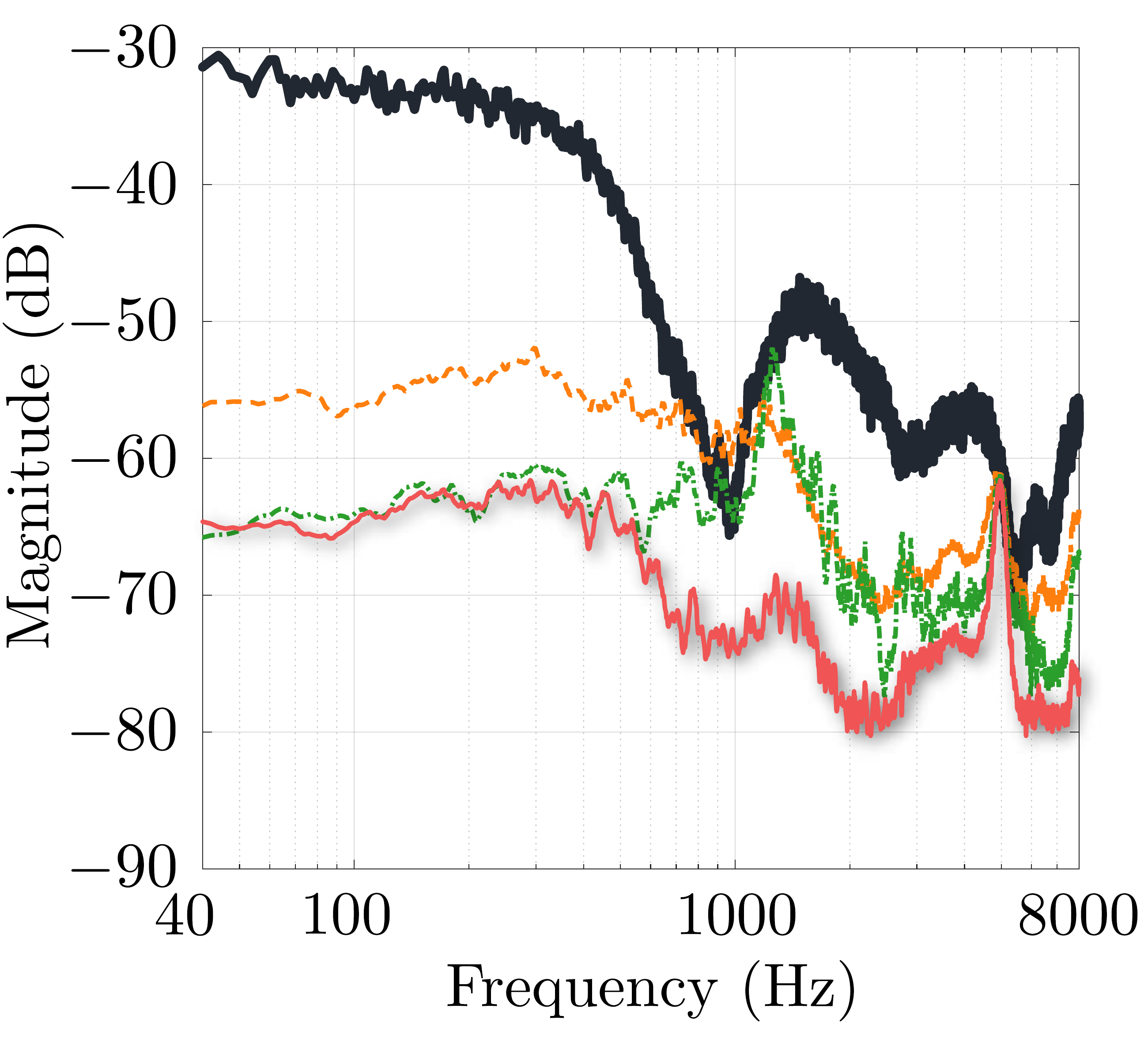}\label{fig:10b}}
        \qquad
        \subfloat[Speech component difference]{\includegraphics[width=1.85in]{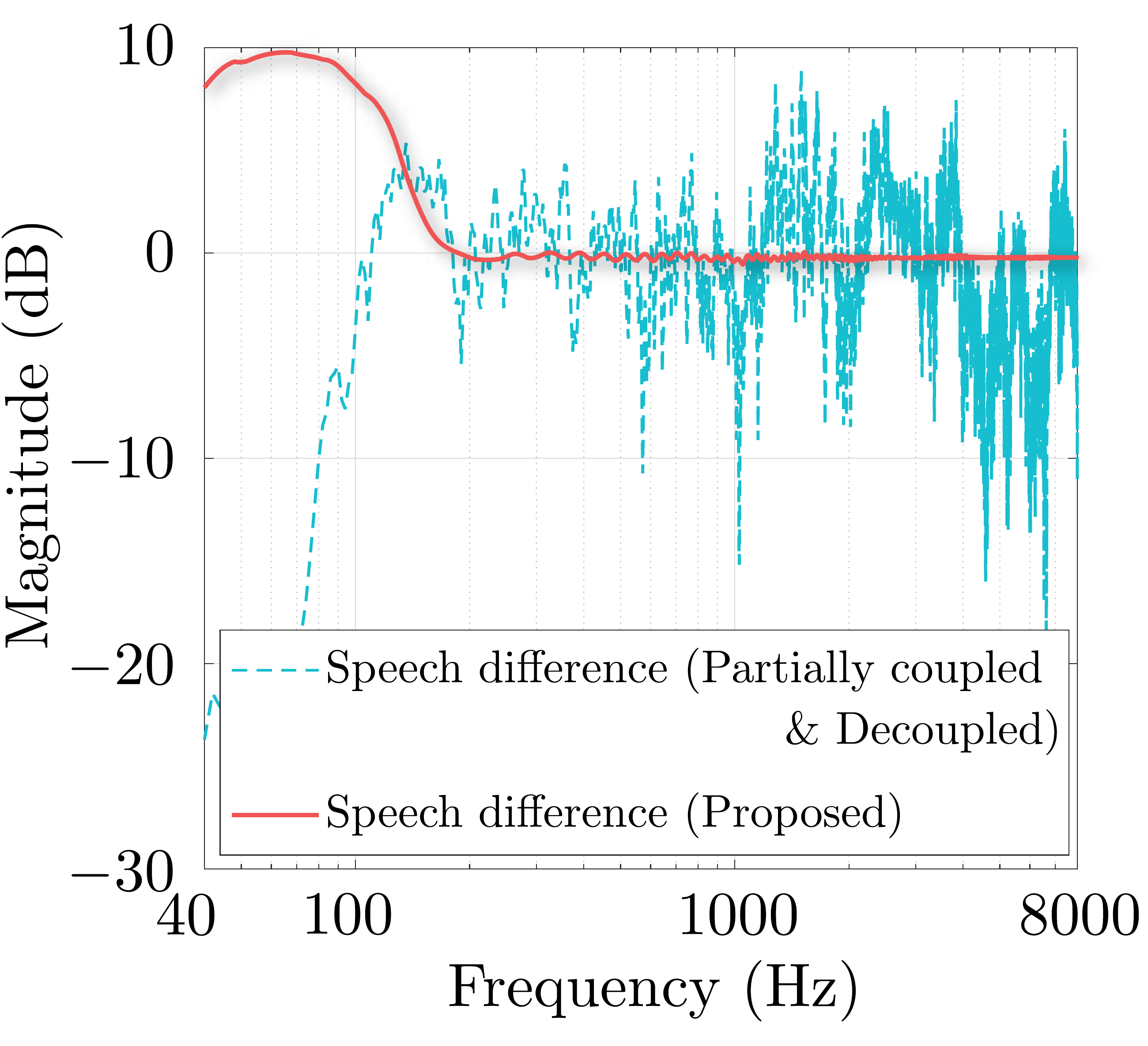}\label{fig:10c}}
        \caption{Spectra of the noisy speech, the clean speech, (a) the overall error signals, (b) the noise components in the error signals and (c) the speech component differences compared to the clean speech in the error signals with ANC enabled in the three configurations. The spectra are from the last 10~s period.}
        \label{fig:10}
    \end{figure*}
    
The ANC performance comparison of the three configurations has been partially demonstrated for a single noise in Fig.~\ref{fig:9c}. A more practical situation is that the noises are from multiple directions. In this case, the uncorrelated pink noise was set coming from five directions $60^\circ, 90^\circ, 120^\circ, 300^\circ$ and $330^\circ$ with the same level, while the DOA of the desired speech remained to be from $0^\circ$ and other configurations were kept the same. Fig.~\ref{fig:10a} shows the overall error signal. The proposed system had the closest result to the desired clean speech. The noise and the speech components in the error signal can be decoupled for further observations.

The noise components in the error signal are shown in Fig.~\ref{fig:10b}. For \replaced{configurations 1 and 2}{the other two configurations}, which tried to cancel both the desired speech and noise, more microphones (in \replaced{configuration 1}{the partially coupled configuration}) can lead to a worse performance. The desired speech in the input signal can result in a greater eigenvalue spread of the correlation matrix of the input signal, thus limiting the step-size and causing a slower adaption speed~\citep{Haykin2002}. The maximum step-size before making the system diverge for \replaced{configuration~1}{the partially coupled system} was 0.00004, whereas the one for \replaced{configuration~2}{the decoupled} was 0.00008. Thus, configuration 2 had a better ANC performance in both Figs.~\ref{fig:9c} and \ref{fig:10b}. The proposed configuration \deleted{3} did not need to cancel the speech. Thus, it has the best NR performance. 

The speech component differences compared to the original clean speech are shown in Fig.~\ref{fig:10c}. \replaced{Configurations 1 and 2}{The other two systems} used the same array for signal extraction, thus having the same result. The reconstructed desired speech from these systems had about $\pm$10~dB difference compared to the clean speech in general. On the other hand, the proposed system did not have this issue since it left the original desired physical sound unaltered. Thus, the difference above 140~Hz was essentially zero. The difference below 140~Hz was due to the spectral weighting filter as discussed previously. This range did not cause any noticeable auditory distortion.

    \begin{figure}[t]
        \centering
        \captionsetup[subfloat]{captionskip=1pt,farskip=1pt}
        \subfloat[Left ear (microphone \#5)]{\includegraphics[width=3.45in]{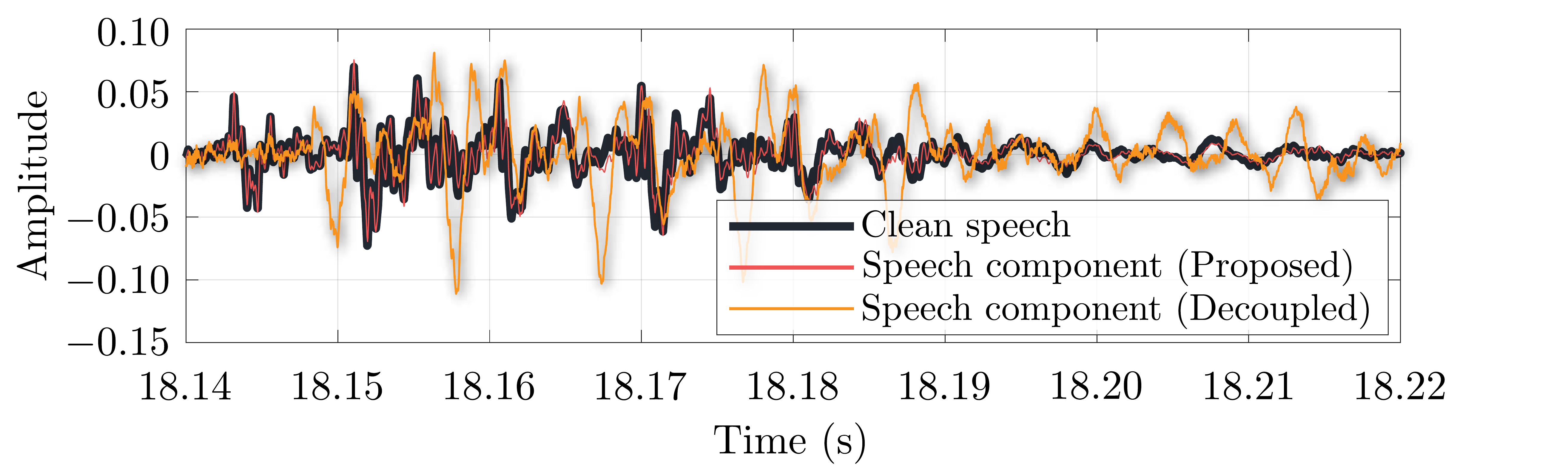}\label{fig:11a}}
        \\
        \subfloat[Right ear (microphone \#6)]{\includegraphics[width=3.45in]{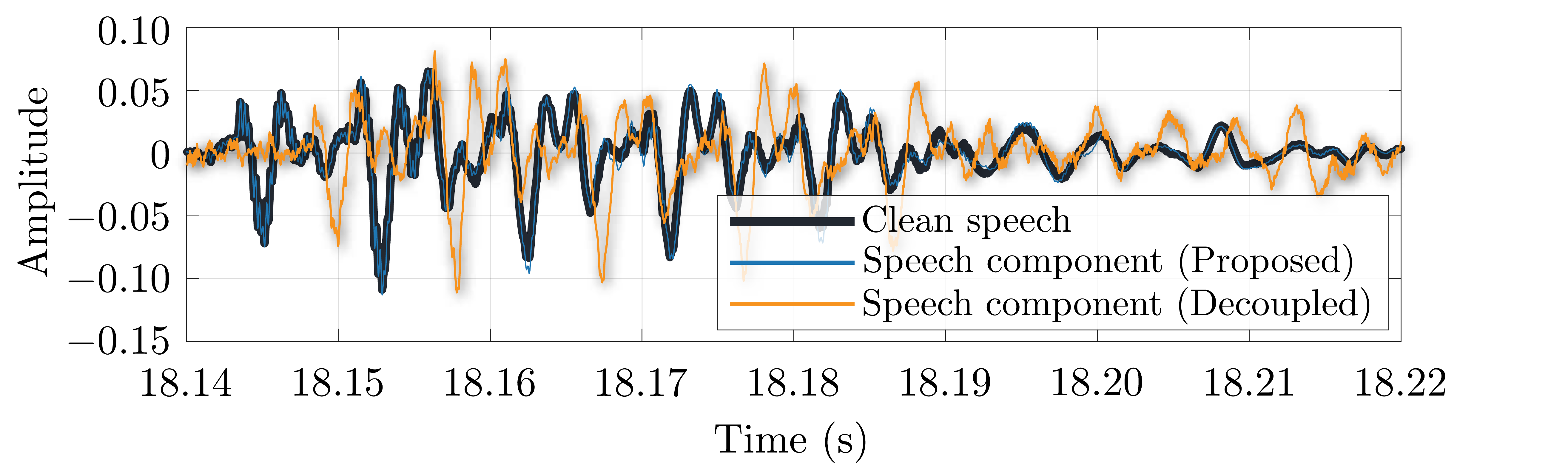}\label{fig:11b}}
        \caption{Waveforms of the clean speech at the error microphone and the speech components in the error signals from the decoupled and the proposed systems at two ears. The desired sound is at $60^\circ$ and the noise comes from $0^\circ$.}
        \label{fig:11}
    \end{figure}
    \begin{figure}[t]
        \centering
        \captionsetup[subfloat]{captionskip=0pt,farskip=0pt}
        \subfloat[Left ear (microphone \#5)]{\includegraphics[width=0.48\columnwidth]{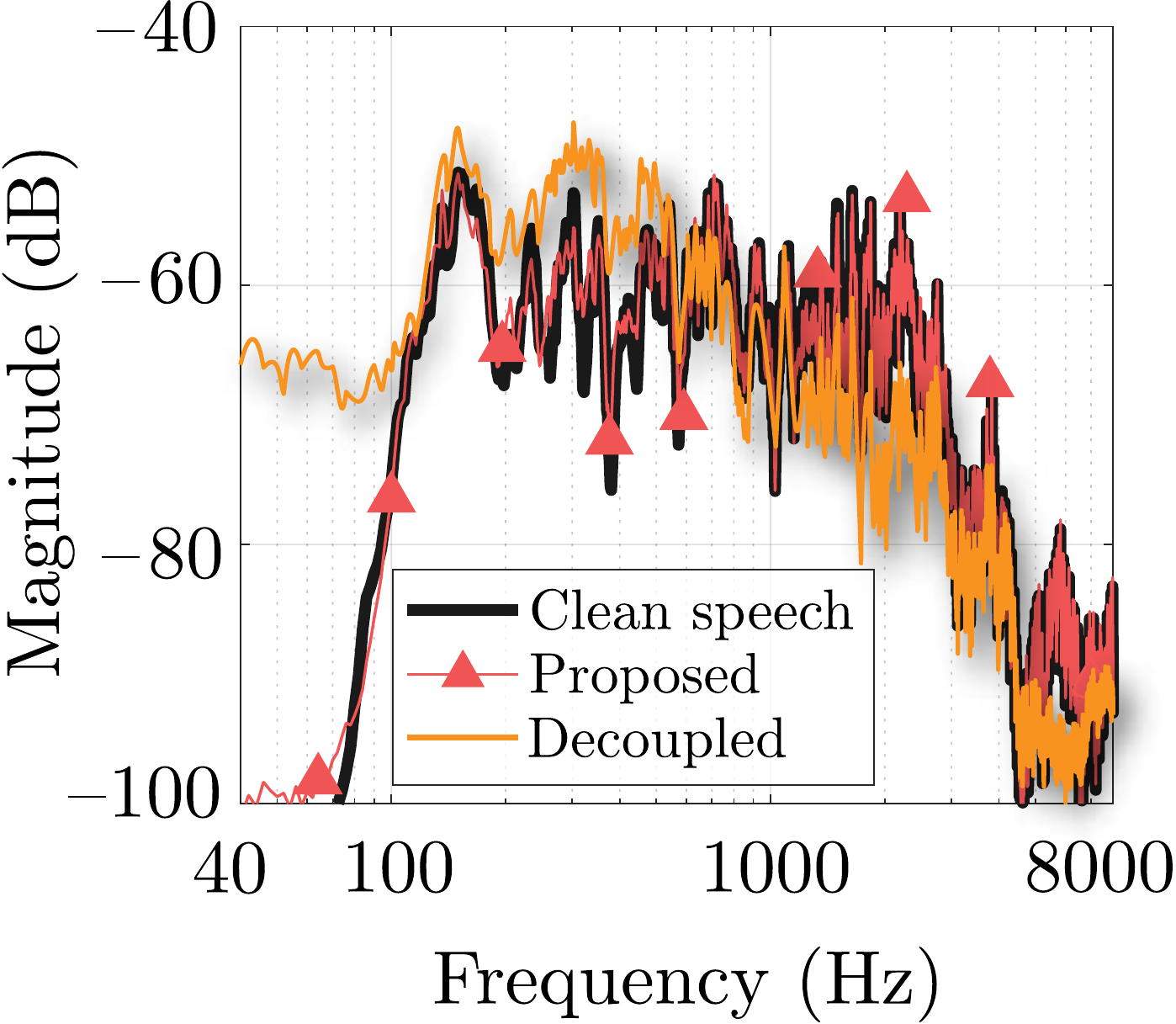}\label{fig:12a}}
        \ \ 
        \subfloat[Right ear (microphone \#6)]{\includegraphics[width=0.48\columnwidth]{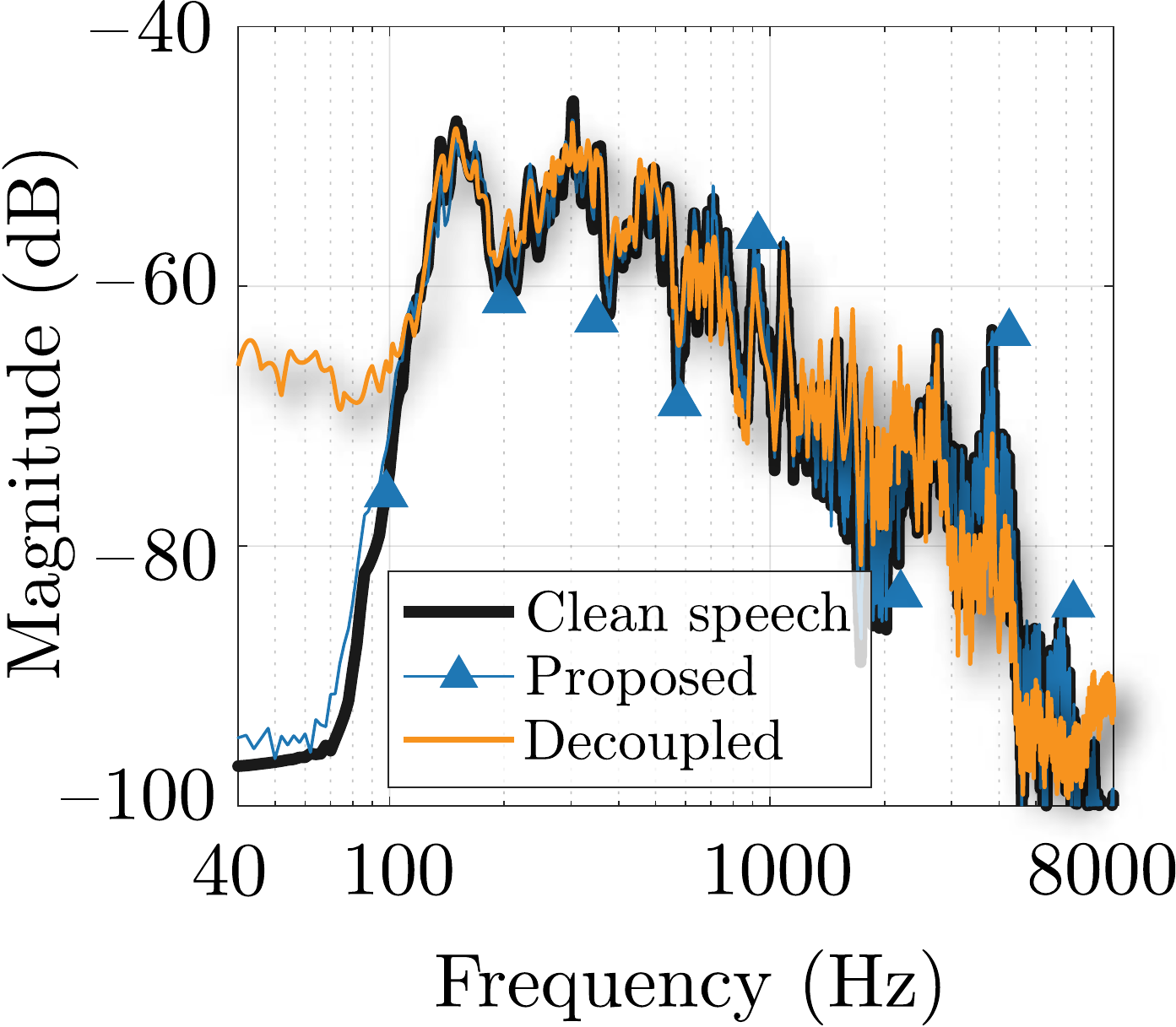}\label{fig:12b}}
        \caption{Spectra of the clean speech at the error microphone and the speech components in the error signals from the decoupled and the proposed systems at two ears. The desired sound is at $60^\circ$ and the noise comes from $0^\circ$.}
        \label{fig:12}
    \end{figure}

\subsection{Binaural localization cues and latency}

When the desired sound comes from other directions than $0^\circ$, binaural localization cues should be enabled for the user. The DOA of the desired speech was moved to $60^\circ$, whereas the pink noise came from $0^\circ$ instead. 

Fig.~\ref{fig:11} shows the waveforms of the speech components in \replaced{configurations 2 and 3}{the decoupled and the proposed systems} compared to the clean speech at two ears. The result in \replaced{configuration 1}{the partially coupled system} was the same as \replaced{configuration 2}{the decoupled} but with 1~ms delay instead of 5~ms. \replaced{Configurations 1 and 2}{The partially coupled and the decoupled systems} had the reconstructed desired signals at microphone \#4. Unless a binaural beamformer was used, the binaural localization cues were lost. \added{This is also shown in Fig.~\ref{fig:12} in the frequency domain. When the DOA of the desired source is at $60^\circ$, the spectral property of the reconstructed desired signal at microphone \#4 may be similar to that of microphone \#6 due to their close proximity, though only between 100~Hz and 2~kHz as shown in Fig.~\ref{fig:12b}. For the left side, the reconstructed signal is monaural and still at microphone \#4, which is significantly different than that of microphone \#5 across the spectrum as shown in Fig.~\ref{fig:12a}.}

On the other hand, it is apparent that the speech component in the error signal from the proposed system agreed with the original clean speech very well, \added{in both time and frequency domains}. This once again confirms that the proposed system can preserve the natural binaural localization cues since it left the desired physical sound unaltered.


\section{Remarks \& Future Directions} \label{Section:remarks}
\added{For the purpose of brevity, it was assumed that there was one desired source throughout this article. For multiple desired sources from different directions, multiple spatial constraints can be added to the cost function in Eq.~(\ref{eq:J_proposed_final}), i.e., $\mathbf{H}_\text{s1}^\text{T}(\tilde{\delta} + {\mathbf{G}} \mathbf{w}) = \mathbf{f}_\text{s1}, \mathbf{H}_\text{s2}^\text{T}(\tilde{\delta} + {\mathbf{G}} \mathbf{w}) = \mathbf{f}_\text{s2}, ...$ pointing to the directions of the desired sources. These constraints can also be combined into one. More information can be found in the linear constraint minimum variance beamforming algorithm~\citep{VanTrees2002,vanveen1988}.  }

\added{The desired sound and the noise can be from the same direction. This can happen when either both sources are located in the same direction, or the noise source is in another direction but the reverberation of the noise is mixed with the desired source. This problem requires further investigation. Possible solutions may be found in the field of \textit{Blind Source Separation}. Studies such as~\citep{mukai2006frequency} can separate mixed signals even when two sources come from the same direction.}

The demonstrated system was a pair of open-fitting AR glasses, which allows the system to be fully coupled, that is, all the microphones to be used for both ANC and the spatial constraint. Another ANC application, for example, a pair of close-fitting ANC headphones, may not be able to use error microphones in the earcups for the spatial constraint since the disturbance signals are highly attenuated. Using all the microphones may not provide any significant improvements. Therefore, using a partially coupled system can reduce computation. 

The secondary sources in such a personal AR glasses system are miniaturized speakers. The proposed system only needs to cancel the noise, so the \replaced{maximum output level requirement for such speakers}{transducer excursion limit} can be relaxed. This \replaced{can be}{is} favorable in acoustic transducer designs. However, the response of miniaturized speakers can still be sub-optimal at some frequencies, which can affect the system control performance. Therefore, the secondary path in this article was from the COMSOL simulation \added{with an ideal sound source to eliminate this factor for the purpose of brevity}. Other applications, such as ANC headrests or ANC windows, \added{which can also adopt the proposed algorithm,} may not have this issue since the loudspeakers are not necessarily required to be miniaturized.

The acoustic feedback from the secondary sources to reference microphones can affect the system significantly in practice. It should be considered in the future.

\section{Conclusion} \label{Section:Conclusion}
In this article, we have presented a spatially selective ANC system that truly preserves the desired sound rather than canceling it and then reconstructing it again. The Frost spatial constraint was imposed on the cost function of a traditional hybrid ANC system, and both the optimal and the adaptive solutions were derived. The method was examined in a pair of AR glasses with six microphones. Overall, the system exhibited good performance in controlling noise coming from undesired directions. The SNR was improved from $-13.9$~dB to 15.2~dB while the SDI was kept at $-25.1$~dB. The system could also maintain good robustness when it was disastrously perturbed. Even when the signal mismatch level was 30~dB higher than that of the desired signal, by choosing the regularization factor based on the largest eigenvalue, the noise could still be controlled by 24.3~dB and the SDI was maintained at $-22.5$~dB. However, the directivity was bound by the causality of the ANC system, which mainly depended on the array configuration. 

Compared to the state-of-the-art systems, the desired sound in the proposed system was indeed preserved while noise from other directions was minimized. The proposed system used much less control effort while still achieving the best ANC performance since it controlled only the noise instead of noise plus speech. Furthermore, when the original desired speech at the error microphone is preserved, there is no need to reconstruct the natural binaural localization cues as in other systems. Future work includes considering acoustic feedback control and examining the system in a non-ideal environment, e.g., in a reverberant room.

\appendix

\section{Optimal solution derivation}\label{Section:appendix_optimal}
By taking the gradient of Eq.~(\ref{eq:J_proposed_final}) and using the chain rule~\citep{Petersen2012}, it is found that
\begin{align}
\nabla_{\mathrm{w}} J&=\frac{\partial J}{\partial \mathbf{w}}
=\frac{\partial \mathbf{u}}{\partial \mathbf{w}}\frac{\partial J}{\partial \mathbf{u}} 
={\mathbf{G}}^\text{T}\left(\Phi_{{\mathbf{x}} {\mathbf{x}}} \mathbf{u}+\mathbf{H} \lambda\right) .
\label{eq:J_nabla}
\end{align}

Setting the gradient of the cost function to zero, the optimal solution can be written as
\begin{align}
    \mathbf{w}_\text{opt} 
    &= - ({\mathbf{G}}^\text{T} \Phi_{{\mathbf{x}} {\mathbf{x}}} {\mathbf{G}} )^{-1} {\mathbf{G}}^\text{T} (\Phi_{{\mathbf{x}} {\mathbf{x}}} \tilde{\delta} + \mathbf{H} \lambda ) \notag
    \\
    &= - \Big( \Phi_{{\mathbf{r}} {\mathbf{r}}}^{-1} \phi_{{\mathbf{r}} {d}} + \Phi_{{\mathbf{r}} {\mathbf{r}}}^{-1} {\mathbf{G}}^\text{T} \mathbf{H} \lambda \Big) ,
    \label{eq:appendix_w_proposed_opt}
\end{align}
where ${\mathbf{r}}(n) = {\mathbf{G}}^\text{T}{\mathbf{x}}(n)$, $\Phi_{{\mathbf{r}} {\mathbf{r}}} = E\left\{{\mathbf{r}}(n) {\mathbf{r}}^\text{T}(n) \right\}$ and $\phi_{{\mathbf{r}} {d}} = E\{{\mathbf{r}}(n) {d}(n) \}$. The Lagrangian $\lambda$ can be found by putting Eq.~(\ref{eq:appendix_w_proposed_opt}) in the spatial constraint $\mathbf{H}^\text{T} ({\mathbf{G}} \mathbf{w}_\text{opt} + \tilde{\delta}) = \mathbf{f}$,
\begin{align}
    \lambda = 
    \Big( \mathbf{H}^\text{T} {\mathbf{G}} \Phi_{{\mathbf{r}} {\mathbf{r}}}^{-1} {\mathbf{G}}^\text{T} \mathbf{H} \Big)^{\dagger} \Big[ (\mathbf{H}^\text{T} \tilde{\delta} - \mathbf{f}) - \mathbf{H}^\text{T} {\mathbf{G}} \Phi_{{\mathbf{r}} {\mathbf{r}}}^{-1}  \phi_{{\mathbf{r}} {d}}  \Big] ,
\label{eq:appendix_lambda}
\end{align}
where superscript $(\cdot )^\dagger$ denotes pseudoinverse. Matrix ${\mathbf{G}}$ is rank deficient due to the delays in the secondary paths in the ANC system. Note that the autocorrelation matrix of the filtered reference signals $\Phi_{{\mathbf{r}} {\mathbf{r}}}$ is typically full rank and thus is invertible.

Using Eqs.~(\ref{eq:appendix_w_proposed_opt}) and (\ref{eq:appendix_lambda}), the optimal solution can be found as
\begin{align}
    \mathbf{w}_\text{opt} = 
    &- \Big[ \textbf{I} - \Phi_{{\mathbf{r}} {\mathbf{r}}}^{-1} {\mathbf{G}}^\text{T} \mathbf{H} \Big(\mathbf{H}^\text{T} {\mathbf{G}} \Phi_{{\mathbf{r}} {\mathbf{r}}}^{-1} {\mathbf{G}}^\text{T} \mathbf{H} \Big)^{\dagger} \mathbf{H}^\text{T} {\mathbf{G}} \Big] \Phi_{{\mathbf{r}} {\mathbf{r}}}^{-1} \phi_{{\mathbf{r}} {d}}  \notag
   \\
   &+ \Phi_{{\mathbf{r}} {\mathbf{r}}}^{-1} {\mathbf{G}}^\text{T} \mathbf{H} \Big( \mathbf{H}^\text{T} {\mathbf{G}} \Phi_{{\mathbf{r}} {\mathbf{r}}}^{-1} {\mathbf{G}}^\text{T} \mathbf{H} \Big)^{\dagger} \Big( \mathbf{f} - \mathbf{H}^\text{T} \tilde{\delta} \Big)
\label{eq:appendix_w_proposed_opt_full}
\end{align}
written in a compact form. Note that $\textbf{f} \neq \mathbf{H}^\text{T} \tilde{\delta}$. Otherwise, it would imply $\textbf{w}=\textbf{0}$ from the cost function in Eq.~(\ref{eq:J_proposed_final}).


\section{Adaptive solution derivation } \label{Section:appendix_adaptive}
The adaptive solution of $\mathbf{w}$ in the proposed method can be found using the LMS method as
\begin{align}
\mathbf{w}(n+1) &=\mathbf{w}(n)-\mu \nabla_{\mathbf{w}} J \notag \\
&=\mathbf{w}(n)-\mu {\mathbf{G}}^\text{T}\left(\Phi_{{\mathbf{x}} {\mathbf{x}}} \mathbf{u}+\mathbf{H} \lambda\right) .
\label{eq:w_n+1_nabla}
\end{align}

Using Eq.~(\ref{eq:u_equals_w}), the constraint has
    \begin{equation}
    \begin{aligned}
    \mathbf{f} &=\mathbf{H}^\text{T} \mathbf{u}(n+1) \\
    &=\mathbf{H}^\text{T}[{\mathbf{G}} \mathbf{w}(n+1)+\tilde{\delta}] \\
    &=\mathbf{H}^\text{T} {\mathbf{G}} \mathbf{w}(n)-\mu \mathbf{H}^\text{T} {\mathbf{G}} {\mathbf{G}}^\text{T}\left[\Phi_{{\mathbf{x}} {\mathbf{x}}} \mathbf{u}+\mathbf{H} \lambda(n)\right]+\mathbf{H}^\text{T} \tilde{\delta} .
    \label{eq:f_lambda}
    \end{aligned}
    \end{equation}

Putting $\lambda(n)$ from Eq.~(\ref{eq:f_lambda}) in Eq.~(\ref{eq:w_n+1_nabla}), it becomes
    \begin{align}
    \mathbf{w}(n+1) 
    =\mathbf{w}(n) 
    &-\mu {\mathbf{G}}^\text{T} \Phi_{{\mathbf{x}} {\mathbf{x}}} \mathbf{u} \notag \\
    &-{\mathbf{G}}^\text{T} \mathbf{H}(\mathbf{H}^\text{T} {\mathbf{G}} {\mathbf{G}}^\text{T} \mathbf{H})^{\dagger}\mathbf{H}^\text{T} {\mathbf{G}} \mathbf{w}(n) \notag \\
    &+\mu {\mathbf{G}}^\text{T} \mathbf{H}(\mathbf{H}^\text{T} {\mathbf{G}} {\mathbf{G}}^\text{T} \mathbf{H})^{\dagger} \mathbf{H}^\text{T} {\mathbf{G}} {\mathbf{G}}^\text{T} \Phi_{{\mathbf{x}} {\mathbf{x}}} \mathbf{u} \notag \\
    &+{\mathbf{G}}^\text{T} \mathbf{H}(\mathbf{H}^\text{T} {\mathbf{G}} {\mathbf{G}}^\text{T} \mathbf{H})^{\dagger}(\mathbf{f}-\mathbf{H}^\text{T} \tilde{\delta}) .
    \label{eq:w_n+1_lambda_n}
    \end{align}

Using Eq.~(\ref{eq:e=d+wGx}) and rearranging Eq.~(\ref{eq:w_n+1_lambda_n}), it becomes
    \begin{subequations}
    \begin{equation} 
    \mathbf{w}(0) = \mathbf{q}, \label{eq:w0_proposed_appendix}  
    \end{equation}
    \begin{equation} \mathbf{w}(n+1) = \mathbf{P} \left[\mathbf{w}(n) - \mu {\mathbf{G}}^\text{T} {\mathbf{x}}(n)e(n) \right] + \mathbf{q}, \label{eq:w_n+1_proposed_appendix} \end{equation}
    \label{eq:w_proposed_appendix}
    \end{subequations}
where
    \begin{subequations}
    \begin{equation} 
    \mathbf{P}=\mathbf{I}-{\mathbf{G}}^\text{T} \mathbf{H}(\mathbf{H}^\text{T} {\mathbf{G}} {\mathbf{G}}^\text{T} \mathbf{H})^{\dagger} \mathbf{H}^\text{T} {\mathbf{G}}, 
    \label{eq:P_proposed} 
    \end{equation}
    \begin{equation}
    \mathbf{q} = {\mathbf{G}}^\text{T}\mathbf{H}(\mathbf{H}^\text{T} {\mathbf{G}} {\mathbf{G}}^\text{T} \mathbf{H})^{\dagger} (\mathbf{f}-\mathbf{H}^\text{T} \tilde{\delta}). 
    \label{eq:q_proposed} 
    \end{equation}
    \label{eq:P_q_proposed}
    \end{subequations}

\section*{Acknowledgment}
The authors wish to thank the anonymous reviewers for their helpful comments and suggestions. The authors would also like to thank Jacob Donley and Sarmad Malik for their assistance with this project.

\bibliography{Reference.bib}

\end{document}